\documentclass[twoside]{aa}  
\usepackage{graphicx}
\usepackage{txfonts}
\usepackage{lscape}

\usepackage{xcolor}

\bibpunct{(}{)}{;}{a}{}{,} 

\begin{document}

   \title{Imaging the warped dusty disk wind environment of SU\,Aurigae with MIRC-X}


   \author{Aaron Labdon
          \inst{1}
          \and
          Stefan Kraus
          \inst{2}
          \and  
          Claire L.\ Davies
          \inst{2}
          \and
          Alexander Kreplin
          \inst{2}
          \and
          Sebastian Zarrilli
          \inst{2}
          \and
          John D.\ Monnier
          \inst{3}
          \and
          Jean-Baptiste le\ Bouquin
          \inst{4}
          \and
          Narsireddy\ Anugu
          \inst{5}
          Benjamin Setterholm
          \inst{3}
          \and
          Tyler Gardner
          \inst{3}
          \and
          Jacob Ennis
          \inst{3}
          \and
          Cyprien Lanthermann
          \inst{5}
          \and
          Theo ten Brummelaar
          \inst{5}
          \and
          Gail Schaefer
          \inst{5}
          \and
          Tim J. Harries
          \inst{2}
          }

   \institute{
   (1) European Southern Observatory, Casilla 19001, Santiago 19, Chile \\
   (2) University of Exeter, School of Physics and Astronomy, Astrophysics Group, Stocker Road, Exeter, EX4 4QL, UK\\
   (3) University of Michigan, Department of Astronomy, S University Ave, Ann Arbor, MI 48109, USA\\
   (4) Institut de Planetologie et d'Astrophysique de Grenoble, Grenoble 38058, France\\
   (5) The CHARA Array of Georgia State University, Mount Wilson Observatory, Mount Wilson, CA 91023, USA\\
   }

   \date{accepted May 18, 2019}

 
  \abstract
   {T Tauri stars are low-mass young stars whose disks provide the setting for planet formation, one of the most fundamental processes in astronomy. Yet the mechanisms of this are still poorly understood. SU\,Aurigae is a widely studied T\,Tauri star and here we present original state-of-the-art interferometric observations with better uv and baseline coverage than previous studies.
   }
   {We aim to investigate the characteristics of the circumstellar material around SU\,Aur, constrain the disk geometry, composition and inner dust rim structure.
   }
   {The MIRC-X instrument at CHARA is a 6 telescope optical beam combiner offering baselines up to 331\,m. We undertook image reconstruction for model-independent analysis, and fitted geometric models such as Gaussian and ring distributions. Additionally, the fitting of radiative transfer models constrains the physical parameters of the disk.
   }
   {Image reconstruction reveals a highly inclined disk with a slight asymmetry consistent with inclination effects obscuring the inner disk rim through absorption of incident star light on the near-side and thermal re-emission/scattering of the far-side. Geometric models find that the underlying brightness distribution is best modelled as a Gaussian with a FWHM of $1.53\pm0.01\,\mathrm{mas}$ at an inclination of $56.9\pm0.4^\circ$ and minor axis position angle of $55.9\pm0.5^\circ$. Radiative transfer modelling shows a flared disk with an inner radius at 0.16\,au which implies a grain size of $0.14\,\mathrm{\mu m}$ assuming astronomical silicates and a scale height of 9.0\,au at 100\,au. In agreement with literature, only the dusty disk wind successfully accounts for the NIR excess by introducing dust above the mid-plane.
   }
   {Our results confirm and provide better constraints than previous inner disk studies of SU\,Aurigae. We confirm the presence of a dusty disk wind in the cicumstellar environment, the strength of which is enhanced by a late infall event which also causes very strong misalignments between the inner and outer disks.
   }


   \maketitle
%

\section{Introduction} \label{sec:intro}
    Outflows from protoplanetary systems are one of the key mass loss mechanisms during the planet formation process. They remove both excess mass and angular momentum from the system, a crucial process as the final masses and rotation rates of stars are known to be significantly less than the inital mass of protostellar cores. 
    
    Within young stellar objects (YSOs) there are several different mechanisms of outflow. Firstly, accretion and magnetospherically driven jets can emerge from the poles of the star. Secondly, photoevaporative winds caused by the ultra-violet (UV) disassociation of molecules in the upper layers of the outer disk can cause significant mass loss. Such photoevaporative winds are usually associated with higher-mass, hotter objects such as Herbig Ae/Be stars. Finally, magnetospherically driven dusty disk winds can originate from the inner disk whereby material is lifted from the disk plane along inclined magnetic field lines. Magnetospheric winds require the presence of a strong magnetic field, usually associated with T Tauri stars with convective envelopes rather than fully radiative interiors. This allows for optically thick material to exist close enough to the central star to contribute to the Near-Infrared (NIR) emission exterior to the main disk structure. This model has been shown to successfully account for the NIR excess of the spectral energy distribution (SED) and the basic visibility features of AB\,Aur, MWC\,275 and RY\,Tau \citep{Konigl11,Petrov19}. While all these mechanisms have been observed, it is not fully understood why some YSOs only appear to exhibit a sub-selection of outflow mechanisms. One of the first stars observed to have an inner dusty disk wind was the T\,Tauri star SU\,Aurigae \citep{Petrov19}. 
    
    For a full description of the literature surrounding SU\,Aurigae and the basic stellar properties, see the previous paper by these authors; \cite{Labdon19} (hence forth LA19).  In this previous work we studied the circumstellar environment of SU\,Aur using interferometric observations from the CHARA/CLIMB and PTI (Palomar Testbed Interferometer) instruments. The disk was found to be inclined at $51.2\pm1.2^\circ$ with a position angle of $61.0\pm1.0^\circ$ and was best modelled with a ring-like geometry with a radius of $0.17\pm0.02\,\mathrm{au}$. Additionally, radiative transfer modelling of visibilities and the SED found that the NIR excess could only be reproduced in the presence of a dusty disk wind, where material is lifted from the disk along magnetic field lines allowing the reprocessing of additional stellar radiation.
    
    \begin{table*}[ht]
        \caption{\label{table:LOG}Observing log from 2018 from the CHARA/MIRC-X interferometer.}
        
        \centering
        \begin{tabular}{c c c c c } 
            \hline
            \noalign{\smallskip}
            Date  &  Beam Combiner &   Stations  & Pointings & Calibrator (UD [mas]) \\ [0.5ex]
            \hline
            \noalign{\smallskip}
            2018-09-13 &  CHARA/MIRC-X   & S1-S2-E1-E2-W1-W2   & 2 & \object{HD 34499} ($0.256\pm0.007$) \\
            
            2018-09-16 &  CHARA/MIRC-X   & S1-S2-E1-E2-W1-W2   & 1 & \object{HD 28855} ($0.303\pm0.008$) \\
            
            2018-09-17 &  CHARA/MIRC-X   & S1-S2-E1-W1-W2   & 2 & \object{HD 40280} ($0.599\pm0.051$) \\
            
            2018-10-26 &  CHARA/MIRC-X   & S1-S2-E1-E2-W1-W2   & 6 & \object{BD+31 600} ($0.391\pm0.011$), \\ 
            & & & & \object{BD+44 1267} ($0.317\pm0.008$), \\
            & & & &\object{BD+43 1350} ($0.318\pm0.008$), \\
            & & & &\object{HD 28855} ($0.303\pm0.008$)\\
            [1ex] 
            \hline 
        \end{tabular}
        \tablefoot{
        All uniform disk (UD) diameters quoted obtained from \citet{Bourges14}.
        }
    \end{table*}

    \begin{figure}[b!]
        \centering
        \includegraphics[scale=0.5]{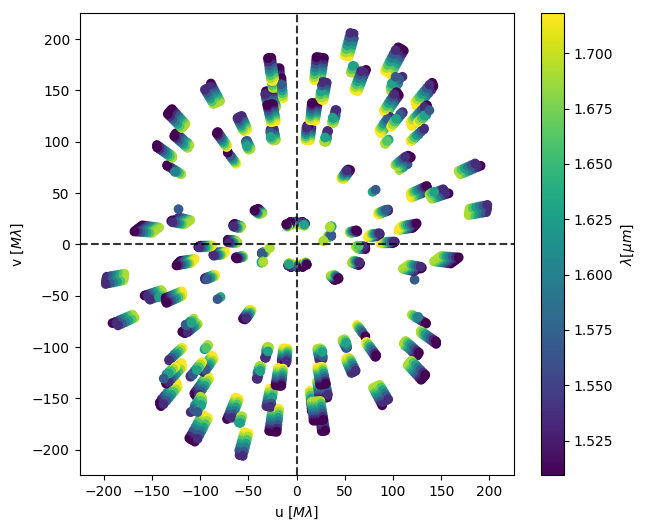}
        \caption{Coverage of the uv plane of the interferometric MIRC-X observations obtained with the CHARA array}
        \label{fig:uvplane}
    \end{figure}

    \begin{figure*}[h!]
        \centering
        \includegraphics[scale=0.45]{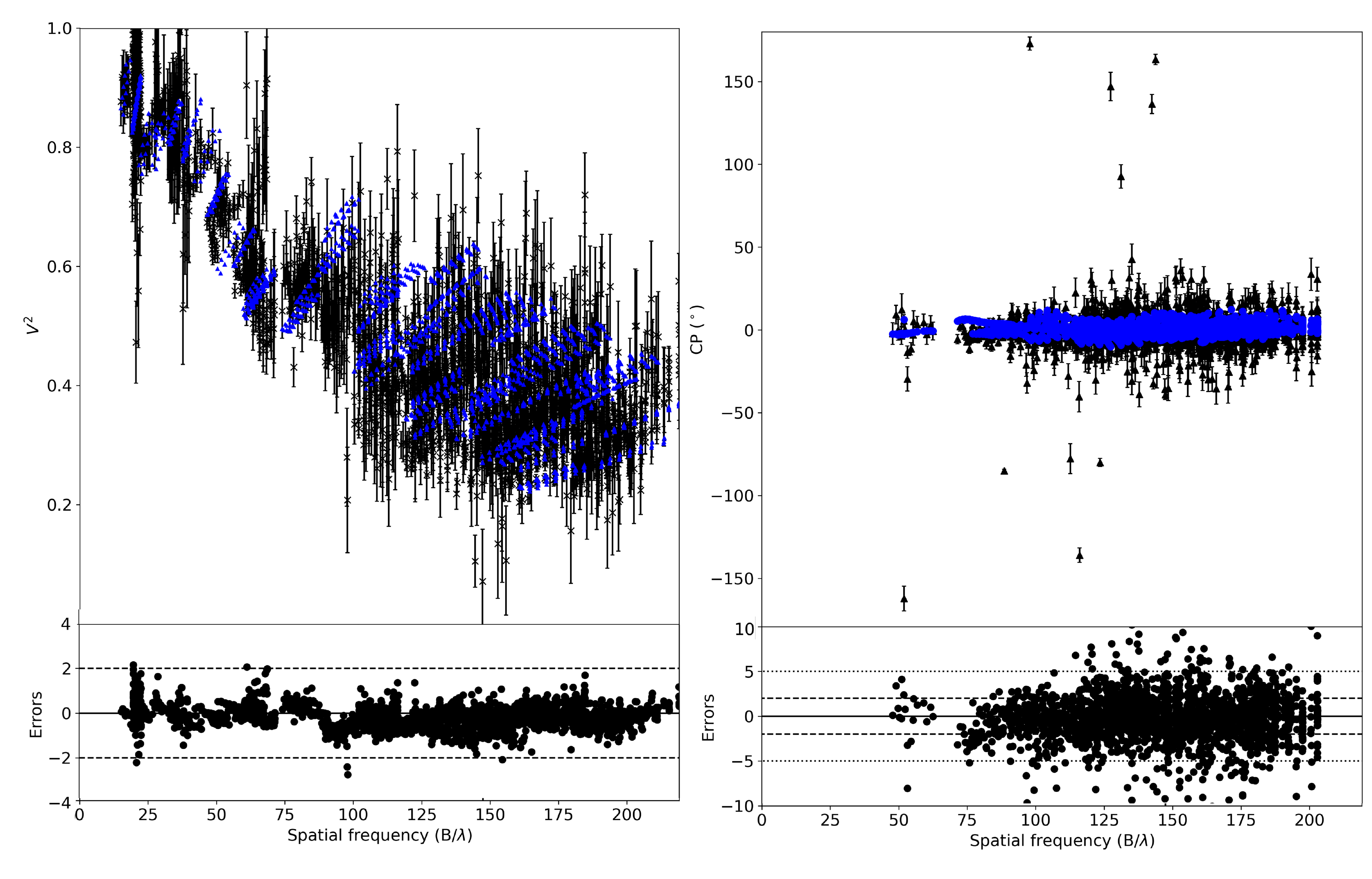}
        \caption{Visibilities and closure phases of the image reconstruction. Black triangles with error bars are the original calibrated observables (squared visibilities on the left and closure phases on right), over plotted as blue markers are the model observables of the reconstructed image. Below each plot is the fit residuals normalised by the standard deviation as black circles. }
        \label{fig:ImRec_V2CP}
    \end{figure*}

    Since the publication of LA19, additional relevant pieces of literature have come to light. Spectroscopic and photometric monitoring of SU\,Aur by \citet{Petrov19} has revealed that a dusty disk wind is the potential source of the photometric variability in both SU\,Aur and RY\,Tau at visible wavelengths. The characteristic time of change in the disk wind outflow velocity and the stellar brightness indicate that the obscuring dust is located close to the sublimation rim of the disk, in agreement with previous theoretical disk wind models \citep{Bans12,Konigl11}. Recent ALMA and SPHERE observations by \citet{Ginski21} reveal a significant disk warp between the inner and out disks of $\sim70 ^\circ$. This misalignment is shown to cause large shadows on the outer disk as it blocks light from the central star. Their observations also suggest that SU\,Aur is currently undergoing a late infall event with significant amounts of material falling inwards from the outermost regions of the disk. Such events have the opportunity to significantly impact the evolution of the disk.
    
    This paper presents one of the first 6-telescope optical interferometric studies of a YSO to date utilising state of the art observations covering a wider range of baseline position angles and lengths (up to 331\,m) (other firsts include \citet{Kraus20,Davies22}). Three different modelling methodologies were used to interpret our data and to provide direct comparisons to LA19. (i)  Image reconstruction was used to obtain a model-independent representation of the data and to derive the basic object morphology. (ii)  Following this geometric model fitting allowed us to gain an appreciation for the viewing geometry of the disk by fitting Gaussian and ring models to the data. In addition, more complex geometric modelling was used to explore the chromaticity of the data. (iii) Finally, we combine interferometry and photometry to derive physical parameters with radiative transfer analysis, where our focus is on confirming the presence of a dusty disk wind.
    
    \section{Observations} \label{Observations}
    
    
    
        The CHARA array is a Y-shaped interferometric facility that comprises six $1\,$m telescopes. It is located at the Mount Wilson Observatory, California, and offers operational baselines between $34$ and $331\,$m \citep{Brummelaar05}. The MIRC-X instrument \citep{Anugu20,Kraus18}, a six-telescope beam combiner, was used to obtain observations in the near-infrared H-band ($\lambda=1.63\,\mu m, \Delta\lambda=0.35\,\mu m$) between September and October 2018. We obtained 11 independent pointings of SU\,Aur, using a mixture of 5 and 6-telescope configurations, due to the short delay line limitations of CHARA. We obtained a maximum physical baseline of $331\,$m corresponding to a resolution of $\lambda/(2B) = 0.70\,\mathrm{mas}$ [milliarcseconds], where $\lambda$ is the observing wavelength and $B$ is the projected baseline. Details of our observations, and the calibrator(s) observed for the target during each observing session, are summarised in Table~\ref{table:LOG}. The uv\,plane coverage that we achieved for the target is displayed in Figure~\ref{fig:uvplane}. Our data covers an exceptionally wide range of baseline lengths and position angles, making the data ideally suited for image reconstruction.

        The MIRC-X data were reduced using the standard python pipeline developed at the University of Michigan by (J.B. le Bouquin, N. Anugu, T. Gardner). The measured visibilities and closure phases were calibrated using interferometric calibrator stars observed alongside the target. Their adopted uniform diameters (UDs) were obtained from JMMC SearchCal \citep{Bonneau06, Bonneau11}, and are listed in Table\,\ref{table:LOG}. 
        
        Considering the short timescale over which the observations were taken the effect of time dependencies/variability of the object is thought to be minimal. However, care was taken to check for time dependencies in the visibilities of baselines of similar length and position angle. Variability in the H band is known to be minimal, so any time dependencies in the visibility amplitudes is likely geometric. However, no significant time dependencies were discovered.

   \begin{figure*}
        \centering
        \includegraphics[scale=0.73]{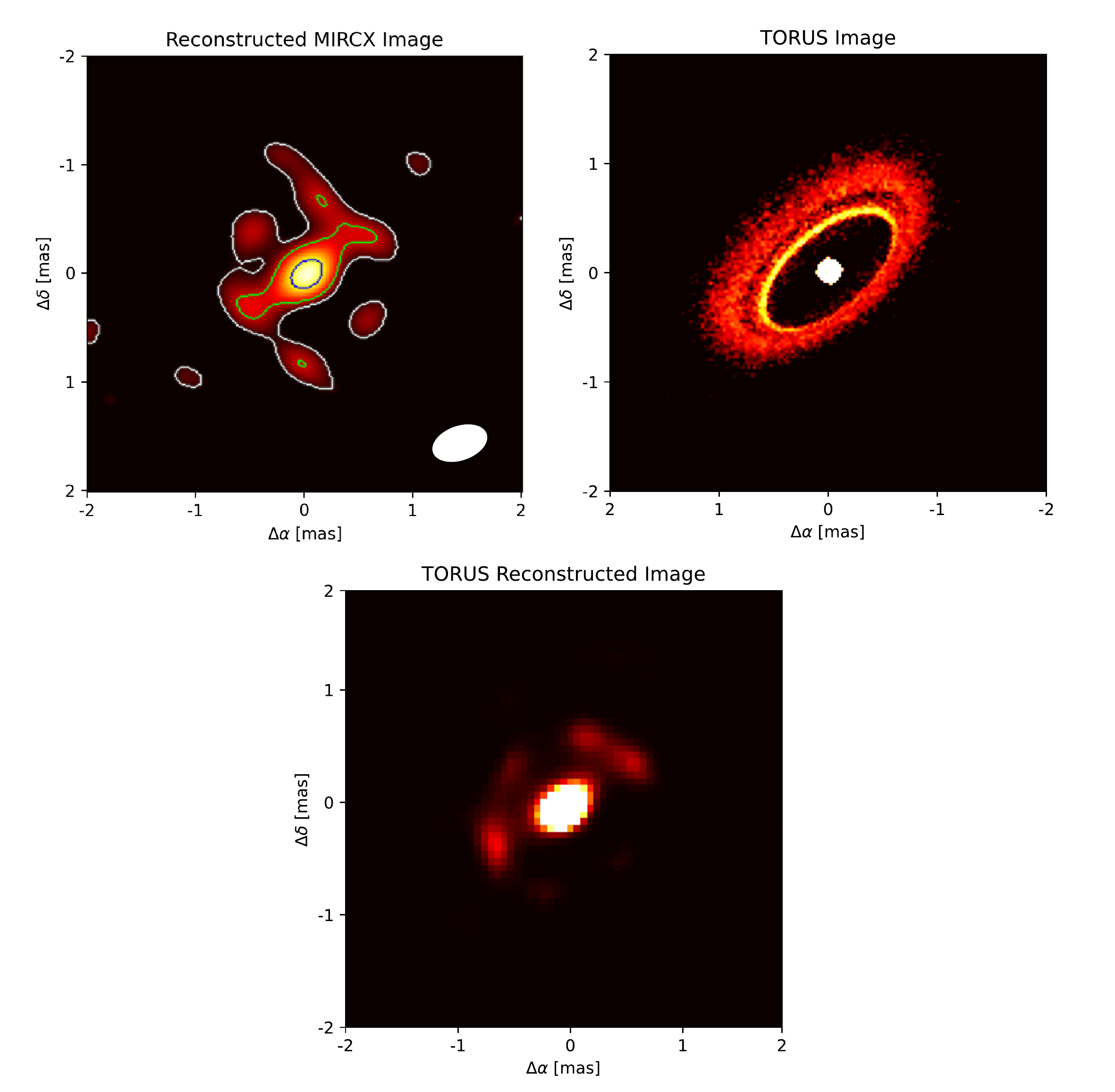}
        \caption{TOP LEFT: Image reconstruction resultant bootstrapped image, including beam size and orientation. The coloured contours represent significance flux levels of 1$\sigma$ (white), 3$\sigma$ (green) and 5$\sigma$ (blue)}.  TOP RIGHT: Radiative transfer image produced using TORUS including a dusty disk wind. BOTTOM MIDDLE: Reconstruction of simulated data created using the best fit TORUS image above, reconstruction parameters are equivalent to those of top left image. Colours are normalised intensity.
        \label{fig:ImRec}
    \end{figure*}

    \begin{figure}
        \centering
        \includegraphics[scale=0.6]{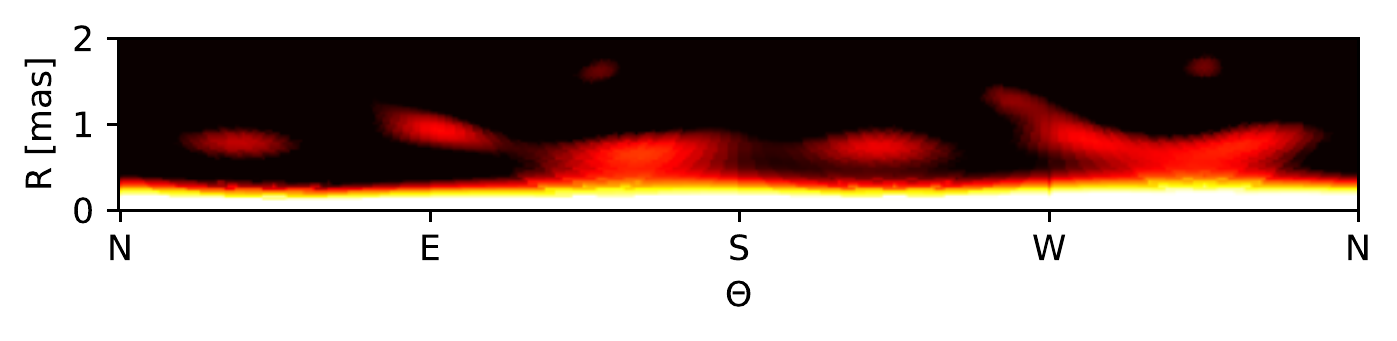}
        \caption{Flattened reconstructed image, created from angular slices though the image from the central star. In the full image of Figure\,\ref{fig:ImRec}, north is up and east is left. R is the radial distance from central star, $\Theta$ is the polar angular direction.}
        \label{fig:FlatImRec}
    \end{figure}

\section{Image Reconstruction} \label{ImRec}
    Image reconstruction techniques require broad and roughly circular uv coverage along as many baseline lengths as possible. Fortunately, the data from these observations lends itself to this process as the uv plane has been well sampled, though some small gaps remain in the position angle coverage. This technique is useful for interpretation of non-zero closure phases, indicative of asymmetric distributions, in a model-independent way. Our closure phase values are shown in Figure\,\ref{fig:ImRec_V2CP}. There are many different algorithms with which to reconstruct images from interferometric data, but the process described here involved the use of the $SQUEEZE$ algorithm \citep{Baron10}. $SQUEEZE$ employs an MCMC approach to image reconstruction and was chosen due to the wide range of available regularisation options and its ability to implement $SPARCO$ a semi-parametric approach for image reconstruction of chromatic objects \citep{Kluska14}. 
    
    
    \begin{table*}[h]
        \caption{\label{table:geomodels}. Best fit parameters for the simple geometric models investigated.
        }
        \centering
        \begin{tabular}{c c c c c} 
            \hline
            \noalign{\smallskip}
            Parameter & Explored Parameter Space & Gaussian & Ring & Skewed Ring    \\ [0.5ex]
            \hline
            \noalign{\smallskip}
            $R$ [mas] & $0.0 - 10.0$ & -- & $0.83\pm0.01$ & $0.17\pm0.16$\\
            $FWHM$ [mas] & $0.0 - 15.0$ & $1.52\pm0.01$ & -- & $0.75\pm0.04$ \\
            $INC$ [$^\circ$] & $0.0 - 90.0$ & $56.9\pm0.4$ & $57.4\pm0.4$ & $56.9\pm0.5$   \\
            $PA$ [$^\circ$] & $0.0 - 360.0$ & $55.9\pm0.5$ & $56.8\pm0.4$ & $55.8\pm0.5$ \\
            $f_{disk}$ & $0.0 - 1.0$ & $0.43\pm0.01$ & $0.32\pm0.01$ & $0.43\pm0.01$ \\
            \hline
            \noalign{\smallskip}
            $\chi^2_{vis}$ & & $11.63$ & $13.87$ & $11.62$ \\
            $\chi^2_{cp}$ & & $6.05$ & $6.05$ & $6.01*$ \\
            \hline
            
        \end{tabular}
         \tablefoot{
        (*) The closure phase quoted is the achieved when allowing the skewed ring to become asymmetric. While the software did detect an asymmetry, it failed to constrain its location in the disk. PA is the minor-axis position angle of the disk and is measured from north ($PA = 0^\circ$) towards east.   
        }
    \end{table*}

    In the $SQUEEZE$/SPARCO routine, the object is modelled as an unresolved central star with an extended, model-independent, environment \citep{Kluska14}. Both components have different spectral behaviours and so differing spectral indices. Additionally, the type and weight of the regularisation was explored, $SQUEEZE$ allows for a very wide range of regularisation algorithms to be implemented. The regularisation plays the role of the missing information by promoting a certain type of morphology in the image. Total variation (TV) was found to most reliably reproduce the best image, TV aims to minimise the total flux gradient of the image and is useful to describe uniform areas with steep but localised changes. These regularisations are considered to be the best ones for optical interferometric image reconstruction \citep{Renard11}. 

    The size and number of pixels also plays an important role in image reconstruction. One cannot simply use the maximum number of pixels of the smallest size to obtain better resolution, they have to be chosen to match uv plane sampling. It was found that a quadratic smoothing regularisation with a weight of $1\times10^5$ and $252\times252$ pixels of $0.1$ mas in size provides the best-fit image reconstruction when utilising exact Fourier transform methods. The optimal regularisation parameters were determined using the L-curve method. 
  
    The final image is shown in Figure\,\ref{fig:ImRec} (top left panel) with the 1,3 and 5-$\sigma$ significance levels shown is white, green and blue contours respectively. The inclination of the disk appears to be greater than that found by LA19 with a similar minor-axis position angle. There also appears to be a central bulge along the minor disk axis likely caused by the over brightness of the star along this axis. The brightness distribution shows a brighter structure along the north-west of the outer disk, parallel to the major axis of the disk. This is consistent with the asymmetry found by LA19 and is indicative of a highly inclined disk where the far side of the inner rim in directly exposed to the observer, while the nearside is obscured by flaring in the outer disk. There are smaller significant structures to the south-east of the disk also, we interpret these to be the shadowed near-side of the rim due to their smaller extent than the northern features. We are not confident in the exact shape of these objects given their irregularity. In order to highlight the radial brightness distribution across the disk, Figure\,\ref{fig:FlatImRec} shows a flattened profile, the elongation of the bulge can be seen in the NW and SE directions with the extended rim material at radii out to 1\,mas.

    In order to quantitatively measure the size and orientation of the emitting region, a simple ellipse was fitted to 3, 4, and 5-$\sigma$ flux significance contours. The averaged results find an inclination of $46^\circ\pm6$ and a position angle of $53^\circ\pm4$. Fitting an ellipse to lower significance levels is not possible due to their irregular shape. the position angle of the ellipses are in good agreement with the values derived in Section\,\ref{GeoMod}. The inclination shows slight deviations from these values.
    
    The chromaticity of the object is measured using two variables in the $SPARCO$ implementation. $f_*^0$, the stellar-to-total-flux ratio at the central wavelength; and $d_{env}$, the spectral index of the extended environment. Only the spectral index of the extended environment is needed as interferometric data is only sensitive to the relative difference in spectral index between the star en circumstellar material. $d_{env}$ is found be $1.7\pm0.8$ which corresponds to a temperature of $1257^{+234}_{-231}$ assuming the objects NIR emission is in a Rayleigh-Jean regime. This is within the range of sublimation temperatures of typical disk astronomical silicates, as expected at such small disk radii. $f_*^0$ is found to be $0.55\pm0.6$, consistent with values measure from the SED and those found from geometric modelling (see Section\,\ref{GeoMod}). The large errors associated with $f_*^0$ and $d_{env}$ are a result of some degeneracy between the parameters, see \citet{Kluska14} for a full description of the procedure.

    
    The visibility and closure phase fits of the image reconstruction are shown in Figure\,\ref{fig:ImRec_V2CP} (top panels) along with the residuals of the fit (bottom panels). The combined visibility and closure phase reduced chi-squared $\chi^2_{red}$ of the image reconstruction was found to be $4.38$.

\section{Geometric Modelling} \label{GeoMod}
    In order to understand the geometry of the system one must consider the application of simple geometric models. In this section we explore several different approaches to modelling our data with both a non-chromatic 'grey' models and techniques which explore the chromaticity. 

\begin{figure}
        \centering
        \includegraphics[scale=0.42]{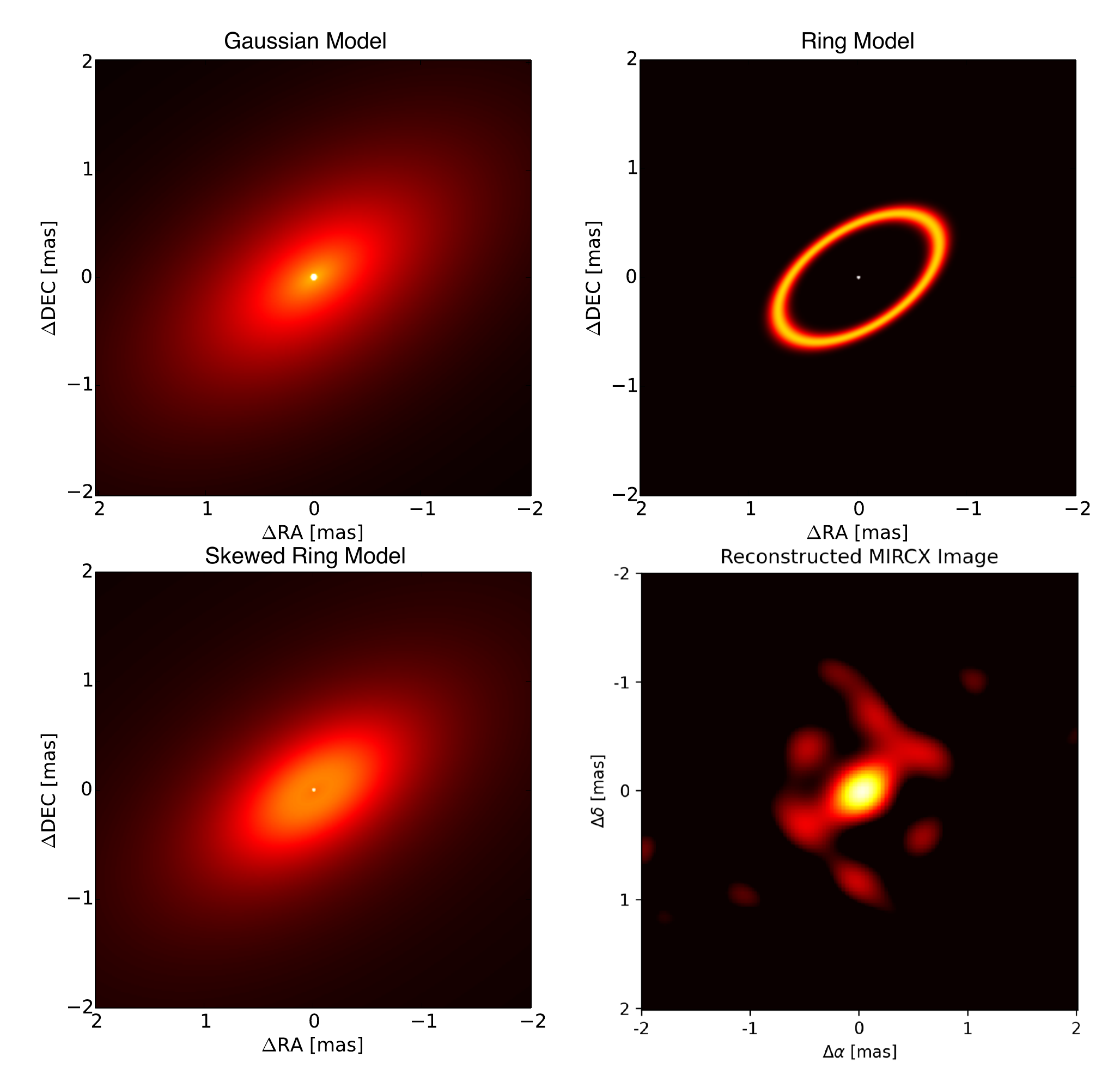}
        \caption{Geometric model images, corresponding to best fit parameters described in Table\,\ref{table:geomodels}. TOP LEFT: Gaussian model brightness distribution. TOP RIGHT: Ring model brightness distribution. BOTTOM LEFT: Skewed Ring model brightness distribution. BOTTOM RIGHT: Full reconstructed image, repeated from Figure\,\ref{ImRec}.}
        \label{fig:ModelImages}
    \end{figure}

    
    \subsection{Basic Geometric Models}
    
    The fitting of Gaussian and ring like distributions to the interferometric variables allows highly accurate estimations of the characteristic size, inclination and position angle of the object. In all models the central star is modelled as a point source, which is an acceptable assumption given the expected angular diameter of the star. The disk parameters are then fitted in the RAPIDO (Radiative transfer and Analytic modelling Pipeline for Interferometric Disk Observations) framework \citep{Kreplin18}, available in-house at University of Exeter. RAPIDO utilises the Markov chain Monte Carlo (MCMC) sampler \emph{emcee} to produce a fit and error estimate \citep{ForemanMackey16}. Three disk models were employed, a standard Gaussian brightness distribution which is characterised by it's full-width-half-maximum (FWHM). Along with two ring models, a sharp ring with a width fixed to 20\% of the disk radius ($R$) and a 'skewed' ring with a more diffuse radial profile produced by convolving with a Gaussian with a $FWHM$. The skewed ring is also capable of modelling azimuthal modulation or disk asymmetries, a detailed description of this model can be found in \citet{Lazareff17}. In addition to the model specific parameters, we also fitted the inclination ($INC$), minor-axis position angle ($PA$) and disk-to-total flux ratio ($f_{disk}$). As we see no evidence of time variability in the data we are able to fit all data simultaneously. The results from the simple geometric model fitting are shown in Table\,\ref{table:geomodels}.

    Out of the geometric models tested, the Gaussian model is considered to be the best fit. Even though the skewed ring produced a slightly small $\chi^2$ value for the closure phase and visibility measurements, we do not consider this significant given the additional degrees of freedom in the model. In addition, as shown in Figure\,\ref{fig:ModelImages}, the best fit skewed ring is tending towards a Gaussian distribution, which a very small inner radius $0.17\,\mathrm{mas}$ with a wide FWHM ring width $0.75\,\mathrm{mas}$. The skewed ring fails to reproduce the small but complex asymmetries seen in the image reconstruction.

    \begin{figure}[t]
        \centering
        \includegraphics[scale=0.5]{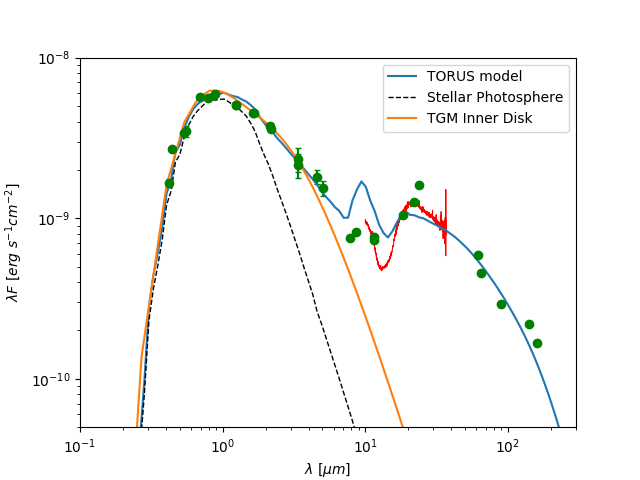}
        \caption{Spectral energy distribution of SU\,Aurigae. Green points are photometric data from a variety of instruments. Red line is Spitzer IR data. Black dashed line is direct radiation from the stellar photosphere. Blue line is the best TORUS computed radiative transfer model inclined at $56^\circ$. Orange line is the SED computed from the simple temperature gradient models described in Section\,\ref{sec:TGMs}.}
        \label{fig:SED}
    \end{figure}

    \begin{figure*}[h]
        \centering
        \includegraphics[scale=0.4]{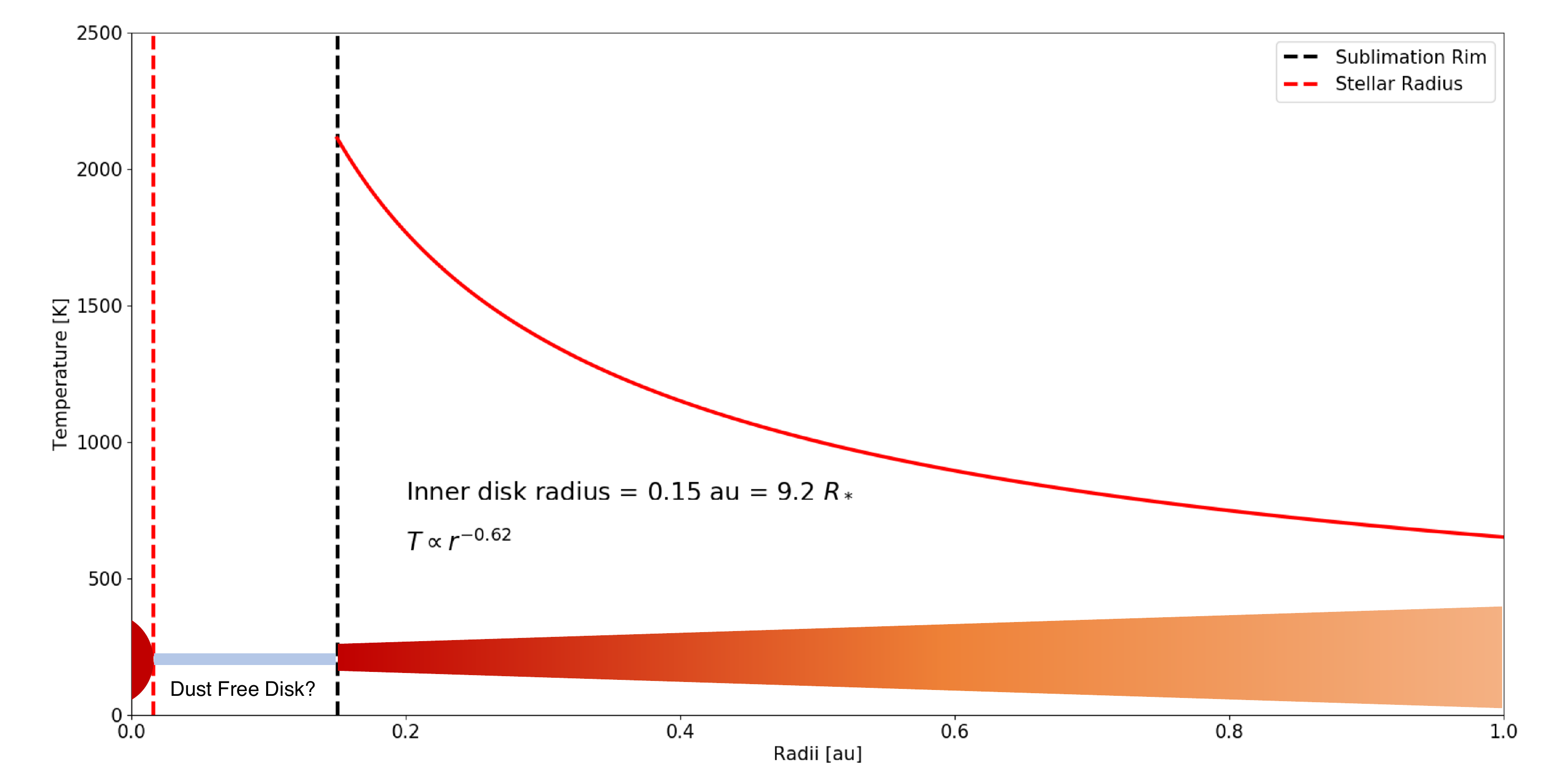}
        \caption{Temperature gradient profile across the inner au of SU\,Aur. The dusty disk extends down an inner radius, consistent with the expected sublimation radius and temperature. Interior to this expected a hot dust free inner disk which is not detected in the continuum observations. }
        \label{fig:TGM}
    \end{figure*}
    
    The best fit Gaussian model finds a disk of FWHM $1.52\pm0.01\,\mathrm{mas}$ which is inclined at $56.9\pm0.4^\circ$ and a minor-axis position angle of $55.9\pm0.5^\circ$. In addition, we find that $43\pm1\%$ of the total flux originates from the disk in the H band. This is consistent with measurements based on the infrared excess of the spectral energy distribution (SED) in LA19 and also with the flux ratio found in the image reconstruction algorithm.
    
    The primary limitation of the simple geometric models described above is that they are intrinsically 'grey' in nature. Meaning they contain no spectral information, hence the large $\chi^2$ values obtained in the fitting process. In order to better model the spectral dependency of the visibility, more complex temperature gradient models that are able to account for observing wavelength must be employed.    

    \subsection{Temperature Gradient Models} \label{sec:TGMs}
    
    A physically correct model can be applied by considering the temperature gradient of the disk.  A temperature gradient model (TGM) allows for the simultaneous fitting of interferometric and photometric observables. The origin of the photometric data used is describe in Table\,\ref{table:Photometry} in Appendix\,A. It is built up by several rings extending from an inner radius $R_{\mathrm{in}}$ to an outer radius $R_{\mathrm{out}}$. Each ring is associated with temperature and hence flux. Therefore, a model SED can be computed by integrating over the resulting blackbody distributions for each of the concentric rings. Such a model allows us to not only to build up a picture of the temperature profile, but also approximate the position of the inner radius. The TGM is based upon a $T_R = T_0(R/R_0)^{-Q}$ profile where $T_0$ is the temperature at the inner radius of the disk $R_0$, and $Q$ is the exponent of the temperature gradient \citep{Kreplin20,Eisner11}. A TGM represents an intrinsically geometrically thin disk. A point source is used at the centre of each model to represent an unresolved star, which is a reasonable approximation given the expected angular diameter of $0.05\,\mathrm{mas}$ \citep{Perez20}. Also included in the fitting of photometric parameters is a treatment of interstellar extinction based on \citep{Fitzpatrick99} with an $E_{(B-V)} = 0.5$ \citep{Bertout07}.
    
    The inclination and position angle of the disk are maintained at fixed values of $56.9\pm0.4^\circ$ and $55.9\pm0.5^\circ$ respectively, from the fitting of the Gaussian distribution. This was done to reduce the number of free parameters in the model. The fitting was undertaken using all of visibility data shown in Figure\,\ref{fig:ImRec} and all the SED points simultaneously. The fitting and error computation was once again done using \emph{emcee} \citep{ForemanMackey16}. 
    
    The results of the temperature gradient modelling are shown in Figure\,\ref{fig:TGM}. We find an inner disk radius of $0.15\pm0.04\,\mathrm{au}$ where the temperature is equivalent to $2100\pm200\,\mathrm{K}$ and decreases with an exponent of $Q=0.62\pm0.02$. The inner disk radius is the point at which the dusty disk is truncated due to the sublimation of material, in contrast to more extreme objects such as FU\,Ori where the inner disk radius is equivalent to that of the stellar radius indicating boundary layer accretion \citep{Labdon21}. Interior to the sublimation radius is expected a hot dust free inner disk from which material can be magnetospherically accreted. However, our low spectral resolution continuum observations are not sensitive to these regions.
    
    An inner disk radii temperature of $2100\pm200\,\mathrm{K}$ may be considered high, but is broadly consistent with laboratory sublimation temperatures for silicate grains. An exponent of $Q=0.62\pm0.02$ is slightly larger than that expected from a disk heated by stellar radiation alone, and may indicate the presence of additional disk heating mechanisms, such as viscous heating \citep{Pringle81,Kenyon87,Dullemond04}.
    
    The resultant SED of the inner disk is shown in Figure\,\ref{fig:SED} as the orange curve. Beyond 5-6\,$\mu m$ the TGM fails to fit the shape of the disk accurately. This is likely due to the flared nature of the disk in contrast to the 'flat' TGM model which will have the strongest effect and longer wavelengths. In addition, the strong disk warp reported by \citet{Ginski21} would also introduce some temperature discontinuity, the modelling of which is beyond the scope of this work.

\section{Radiative Transfer} \label{RadTrans}

    In order to provide a more physical model and directly compare to LA19, we used the TORUS Monte-Carlo radiative transfer code \citep{Harries19} to simultaneously fit the visibility, closure phase and photometric data of the SU\,Aurigae system.
    
    The models adopted here are based on the disk models used by LA19, adapted to account for the higher inclination and different observing wavelength. In these TORUS simulations, the dust was allowed to vertically settle to the scale height of the gas component and the dust sublimation radius was left as a free parameter, allowing the inner rim radius to define itself based on well-defined rules of the \citet{Lucy99} iterative method to determine the location and the temperature structure of the whole disk. This is implemented whereby the temperature is initially calculated for grid cells in an optically thin disk structure, with dust added iteratively to each cell with a temperature lower than that of sublimation, until the appropriate dust to gas ratio is reached ($0.01$). We confirmed that stellar photosphere models of \citet{Kurucz04} using these stellar parameters can reproduce the photometry measurements of SU\,Aur reasonably well across the visible continuum. We adopt a silicate grain species with dust properties and opacities adopted from \citet{Draine03}. For a more detailed description of TORUS and the algorithms used, see \citet{Davies18,Labdon19}.
    
      \begin{sidewaystable}
        \caption{\label{table:torus}Best-fit parameters resulting from TORUS \citep{Harries00} radiative transfer SED and visibility modelling.
        }
        \centering
        \begin{tabular}{c c c c c c} 
            \hline
            \noalign{\smallskip}
            Parameter  & Literature value & Reference & Range explored &  Best fit value & Notes   \\ [0.5ex]
            \hline
            \noalign{\smallskip}
            $R_\mathrm{{inner}}$ & $0.18~\mathrm{au}$ & (1)(2) & $0.1 - 0.6~\mathrm{au}$ & $0.16~\mathrm{au}$ & Determined based on the gas density and grain size\\
            $R_\mathrm{{outer}}$ & $100~\mathrm{au}$ & (1)(3) & - & $100.0~\mathrm{au}$ & Fixed to literature values    \\
            $h_\mathrm{0}$ & $15.0~\mathrm{au}$ & (3) & $7.0 - 20.0~\mathrm{au}$ & $9.0~\mathrm{au}$ & Scale height at 100\,au    \\
            $\alpha_\mathrm{{disk}}$ & $2.4$ & (3) & $1.0 - 3.0$ & $2.3$ & $\mathrm{\alpha_{disk}\,\,is\,\,fixed\,\,at\,\,(\beta_{disk}+1)}$ \\
            $\beta_\mathrm{{disk}}$ & $1.4$ & (3) & $0.0 - 2.0$ & $1.3$  \\
            $\mathrm{Dust:Gas}$ & $0.01$ & (1)(3) & - & $0.01$ & Fixed to literature values  \\
            $a$ & $0.14\,\mathrm{\mu m}$ & (1) & $0.1-1.4~\mathrm{\mu m}$ & $0.14~\mathrm{\mu m}$ \\  
            $T_\mathrm{{sub}}$ & $2000~\mathrm{K}$ & (4) & - & $2000~\mathrm{K}$ & Fixed to literature values  \\ [1ex] 
            \hline
            \noalign{\smallskip}
            Dusty disk wind parameter & Literature value & Reference & Range explored & Best fit value \\
            \hline
            \noalign{\smallskip}
            $R_\mathrm{{0min}}$ & $4.5$ & (3) &  2.0-6.0 & $4.5~\mathrm{R_{\sun}}$ & New grid of models\\
            $T_\mathrm{{wind}~(near~surface)}$ & $1600~\mathrm{K}$ & (3) & 1200-2400 & $1600~\mathrm{K}$ & New grid of models\\
            $\mathrm{Opening~Angle}$ & $50^\circ$ & (3) & $25 - 65^\circ$ & $45^\circ$ & New grid of models\\
            $\dot{M}$ & $10^{-7}~M_{\odot}yr^{-1}$& (3) & $10^{-10}$ - $10^{-6}$ & $10^{-7}~M_{\odot}yr^{-1}$ & New grid of models\\
            \hline
            
        \end{tabular}
        \tablebib{
        (1) \citet{Akeson05}; (2)~\citet{Jeffers14}; (3)~\citet{Labdon19}; (4) \citet{Pollack94}
        }
    \end{sidewaystable}


    The dusty disk wind model is adapted from \citet{Bans12}. This mechanism is based on the presence of a large-scale, ordered magnetic field which threads the disk along which disk material is flung out. The high magnetic pressure gradient above the disk surface accelerates the material which is then collimated through the azimuthal and poloidal field components \citep{Bans12}. These centrifugally driven winds are highly efficient at distributing density above and below the plane of the disk, carrying angular momentum away from the disk surface. A full description of the implementation within the TORUS radiative transfer code can be found in LA19.

    The disk model adopted follows the curved inner rim prescription of \citet{Isella05} with a density dependent sublimation radius whereby grains located in the disk midplane are better shielded due to higher densities and so can exist closer to the central star than grains in the less dense upper layers. A full summary of the disk parameters can be found in Table\,\ref{table:torus}. The only parameters we explored with respect to LA19 were the disk scale height ($h_0$), the grain size ($a$) and flaring index $\beta_{disk}$. The key difference in the models described here compared to LA19 is the grain size adopted. Here we adopt a smaller grain size of $0.14\,\mu m$, which in turn leads to a slightly smaller inner radius of $0.16\,\mathrm{au}$. Additionally, in order to improve the SED fit at longer wavelengths we also adopt a more modest scale height of $9.0\,\mathrm{au}$. The resultant SED from the radiative transfer model is shown in figure\,\ref{fig:SED}, in addition to the model stellar photosphere, also calculated within TORUS. The SED is fit well across the optical and IR and longer mm wavelengths, however is a relatively poor fit across the 8-40\,$\mu m$ range. We attribute this to the disk warp reported in \citet{Ginski21} which would result in a physical disk break and temperature discontinuity. The geometry governing disk warps is little understood, in particular how it would effect the inner and outer edge of the warp. The shapes of these rims, and how is the vertical structure of the disk effected at this point, are not known. Such a model are well beyond the scope of this paper for 2 reasons: Firstly we lack high spatial resolution data at M/F-IR wavelengths which might cover the location of the disk warp. The complex geometry of the warp would require extensive modelling, which is difficult given the limitations of the radiative transfer code used.

    \begin{figure*}
        \centering
        \includegraphics[scale=0.56]{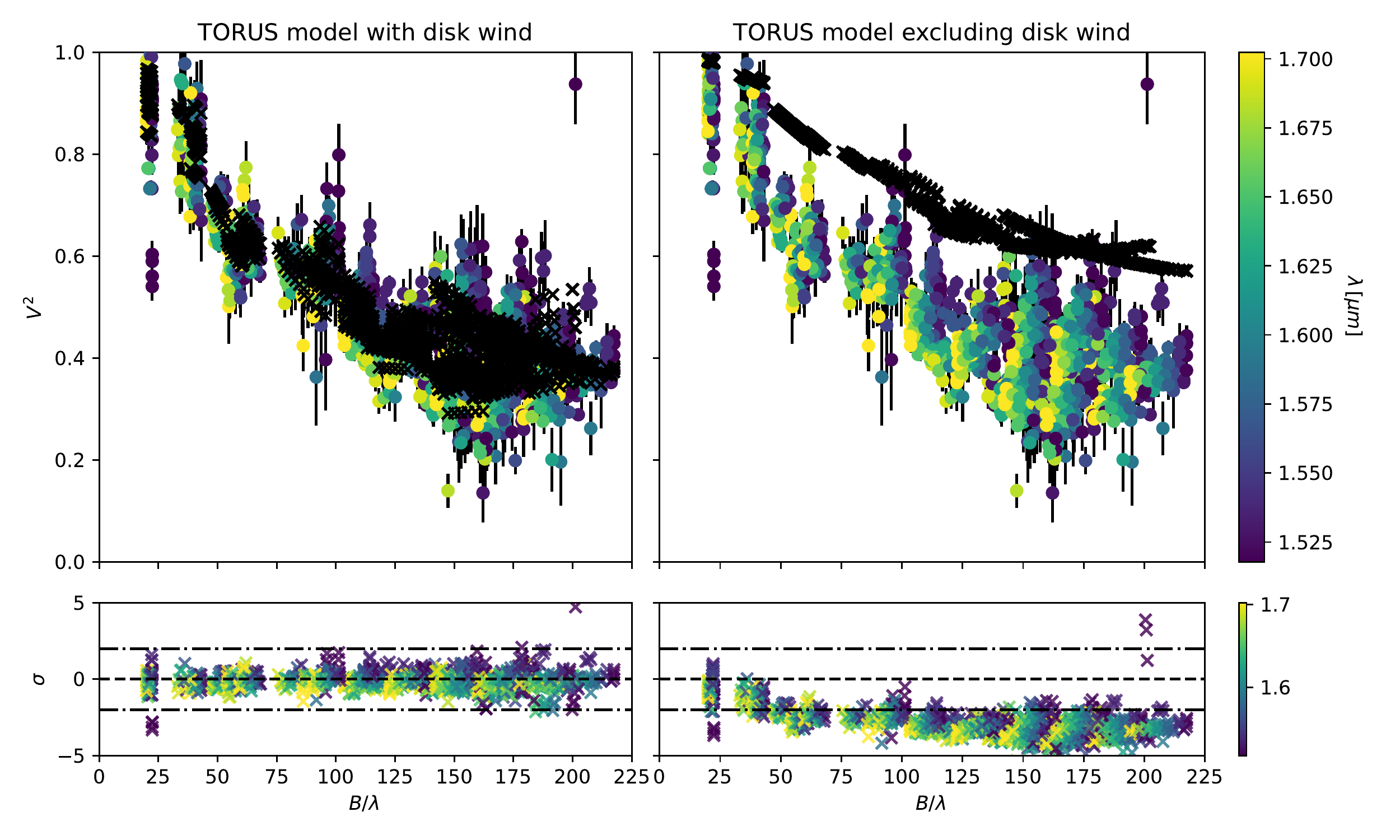}
        \caption{Coloured circles are the MIRC-X squared visibilitites, where the colour represents the wavelength across the H-band. Black crosses are the corresponding squared visibilities extracted from TORUS radiative transfer model images. LEFT: Best fit TORUS model including a strong dusty disk wind component. RIGHT: The same base disk model, excluding any dusty disk wind component. Lower panels show the residuals in the fits between data and models.}
        \label{fig:Vis_TORUS}
    \end{figure*}
    
    The presence of a disk wind is once again required to fit both the visibilities and the SED. The absence of a disk wind fails to reproduce IR excess across both the H and K bands, with insufficient NIR disk flux. This significantly impacts both the SED and interferometric fit. A disk wind is required to eject more hot dust above the midplane of the disk where it is directly exposed to stellar radiation which is reprocessed as an IR excess. 
    
    This is shown in Figure\,\ref{fig:Vis_TORUS}, where the squared visibilities are shown for both a disk model with and without the dusty wind environment. The disk wind model provides a far superior fit to the observations, being able to successfully reproduce the NIR excess. Beginning from the wind parameters found in LA19, we run a new grid of models exploring the disk wind parameter space. The parameters explored and their ranges are described in Table\,\ref{table:torus}, along with the resulting best fit values. The results are broadly similar to those found in LA19, with exception of a slightly higher temperature of material in the disk wind, closer to dust temperatures found in the inner disk. Of particular note is the into-wind accretion rate of $10^{-7}\,\odot{M}$. Considering the historically accepted on-to-star accretion rate to into-wind accretion ratio of $0.1$, this level of transport is perhaps unphysically high given the age of the star, but a discussion of possible mechanisms is included in Section\,\ref{Discuss}. 
    

    The final computed image is shown in Figure\,\ref{fig:ImRec} (middle left panel) and shows the clear asymmetry originating from the inclination in the asymmetry map in the same figure (middle left panel). In order to approximate what this computed image would look like if observed at the same resolution as the original observations, we computed synthetic visibility and closure phases based on the radiative transfer images (as described in \citep{Davies18}). Artificial noise and error bars were computed to be representative of the original data and to ensure an accurate representation. These synthetic observables were then reconstructed in the same manner as the original data, as described in Section\,\ref{ImRec}. Care was taken to ensure the consistency in the reconstruction parameters for both the real and synthetic observables. The reconstructed TORUS image is also shown in Figure\,\ref{fig:ImRec} (bottom right panel), and shows clear similarities with both the original TORUS image and the image reconstructed from the original data.

\section{Discussion} \label{Discuss}

    Our extensive observations and analysis of the circumstellar environment of SU\,Aurigae have revealed the details of the inner disk in unprecedented detail. The wide variety of techniques used to analyse our interferometric data allows us to precisely define the disk characteristics. 

    Image reconstruction is a crucial, model independent, method of analysis which is ideally suited to our dataset with extensive uv and baseline coverage. Our analysis reveals an elliptical shape, indicative of an object with a high inclination as shown in Figure\,\ref{fig:ImRec}. There appears to be a central bulge to the disk, however this feature is not thought to be physical but rather a manifestation of the brightness of the central star combined with the width of the disk at this point. The more extended material in the image are thought to be a depiction of the far-side of the disk rim which is un-obscured by the outer disk, this re-enforced by the fact that the extended material is bright in the north-east of the image. There is a significant asymmetric feature in the form of a thin brightness on the north-eastern edge of the disk. The unique shape of this feature indicates that this is again caused by the high inclination obscuring the nearside disk rim. The effect of an inclined disk on the observed brightness distribution is described extensively by \cite{Jang13} and accurately describes the observations here. By fitting an ellipse to the significant features in the image, we gain a quantitative measure of the disk geometry. The results find an inclination of $46^\circ\pm6$ and a position angle of $53^\circ\pm4$, very similar to those derived from more reliable geometric model fits described below. The scale and shape of the image is similar to that of LA19, with a slightly higher inclined viewing angle. We consider the images of this work to be the more accurate depiction of SU\,Aur given the higher quality observations, taken over a much shorter timescale, with the added detail and resolution this entails.
    
    Although geometric modelling is much more constrained in the geometries it can explore, in does provide a more quantitative view of the disk. It was found that the model which best fit our data was a simple Gaussian distribution with a point source representing the star. A Gaussian model is consistent with other work, both on this object by LA19, but also in other YSO studies such as the survey by \citet{Lazareff17} who find that little under half of their 51 objects can be modelled by a Gaussian structure. The Gaussian fitted in this work has a FWHM of $1.52\pm0.01\,\mathrm{mas}$ ($0.239\pm0.002\,\mathrm{au}$) at an inclination of $56.9^\circ\pm0.4$ at a minor axis position angle of $55.9^\circ\pm0.5$ with a stellar-to-total flux ratio of $0.57\pm0.01$. The reduced $\chi^2$ value for the visibilities is $11.63$ and $6.05$ for the closure phases, which are equivalent to $0^\circ$ for this centro-symmetric model. These values are in agreement with the literature values of \citet{Akeson05} who find a K band radius of $0.18\pm0.04$\,au and an inclination of $62^{\circ+4}_{-8}$. Similar values for the inclination in literature are \textasciitilde$60^\circ$ and \textasciitilde$50^\circ$ found by \citet{Unruh04,Jeffers14} respectively. The minor axis position angle derived here is significantly greater than the literature values of $24^\circ\pm23$ and $15^\circ\pm5$ found by \citet{Akeson05,Jeffers14}. This difference is likely due to either: The poor uv coverage and lack of longer baselines in previous interferometric studies, both of which make estimating the position angle and inclination particularly unreliable. Other non-interferometric studies focus on the outer disk which is shown by \citet{Ginski21} to be misaligned compared to the inner disk. 
    
    The geometric modelling results are broadly similar to those presented in our previous work LA19 where an inclination of $50.9\pm1.0^\circ$ and minor axis position angle of $60.8\pm1.2^\circ$ were found and the data were marginally better described by a ring-like brightness distribution. The values and models presented here are considered to be more accurate due high precision observations and significantly smaller potential for temporal variations, as the data of LA19 was coalesced over 14 years. These values are also consistent with observations of the outer disk by \citet{Ginski21} where dark shadows are observed in scattered light origination from a significant disk warp between the inner and outer regions. On larger scales the near-side of disk is seen to the north-east, in our observations it is seen to the south west. 
    
    \begin{figure}
        \centering
        \includegraphics[scale=0.45]{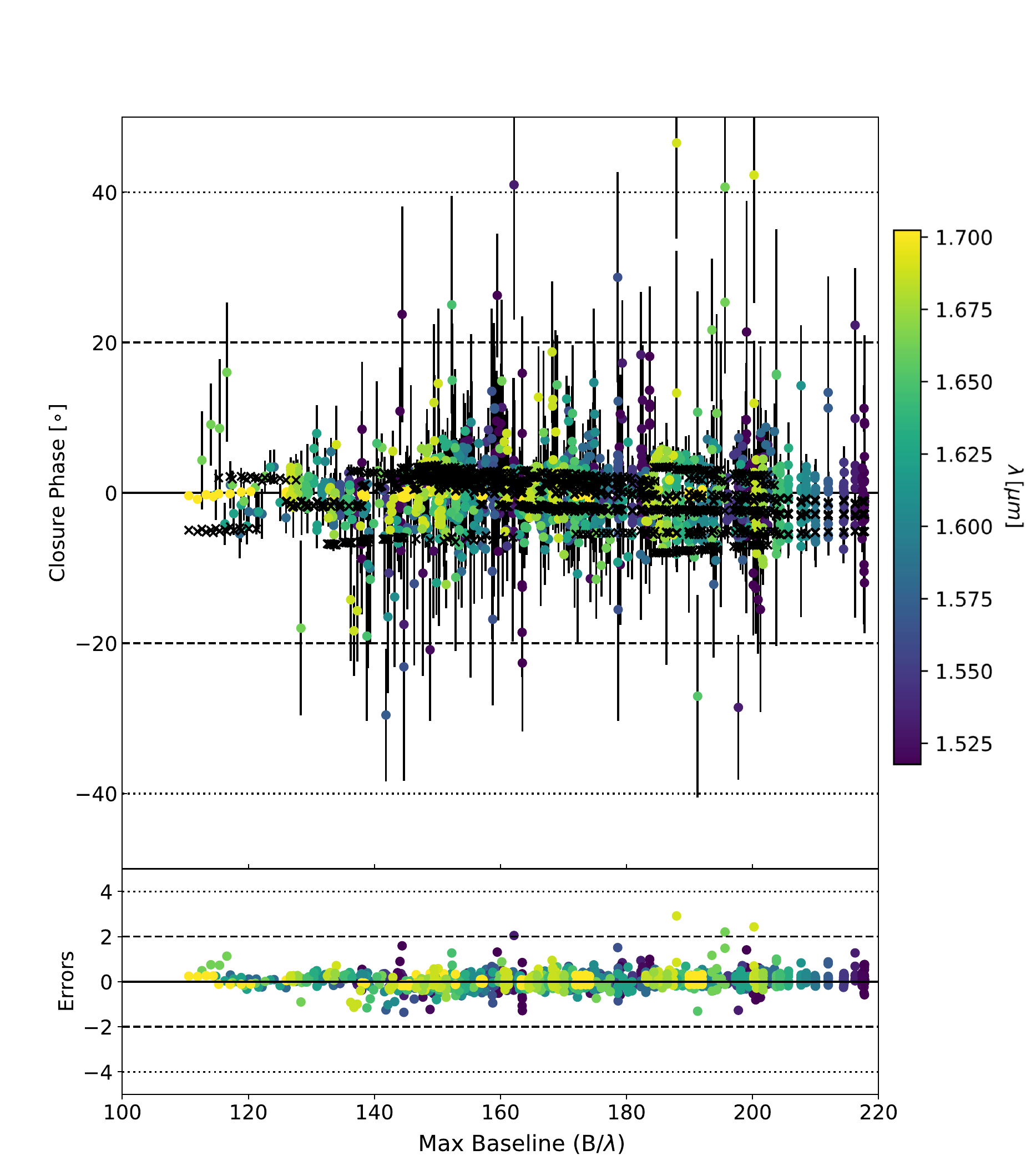}
        \caption{Black crosses are the closure phase data obtained with MIRC-X. Overlaid as coloured points are the TORUS model closure phases, the colours represent the wavelength of the spectral channels and follows the same convention as other plots in this work. Below in black points are the normalised residual errors of the fit. }
        \label{fig:Closure_TORUS}
    \end{figure}
    
    In modelling the temperature gradient of SU\,Aur we can gain an appreciation for spectral dependence of our interferometric variables across the 6 spectral channels of MIRC-X. Our modelling finds a disk which extends down to $0.15\pm0.04\mathrm{au}$ where the temperature is equivalent to $2100\pm200\,K$ and decreases with an exponent of $Q = 0.62\pm0.02$. The outer edge of this temperature regime was found to be $0.20\pm0.03\mathrm{au}$, showing this prescription only covers the innermost regions of the disk. Modelling outer regions of the disk is beyond the scope of this paper, as our NIR interferometric data does not cover emission from these regions and larger radii are likely effected by the significant disk warp causing a temperature gradient discontinuity. A temperature gradient exponent of $Q = 0.62\pm0.02$ found here lies between two established models from literature. That of \citet{Pringle81} who find that a steady state, optically-thick accretion disk heated by viscous processes will exhibit an exponent of $0.75$ and of \citet{Kenyon87,Dullemond04} who show that a flared disk heated by reprocessed stellar radiation alone will exhibit an exponent of $\le 0.5$. As such, it is difficult to comment on the heating of the circumstellar environment of SU\,Aurigae. But it is possible that the disk is not heated by stellar radiation alone, but additional heating processes such as viscous heating may also be present.
    
    The radiative transfer modelling presented in this paper is heavily based on LA19. For a detailed discussion of the motivation behind certain choices, particularly in relation to the shape of the inner rim, we recommend the reader see Section\,5 of that paper. In this work we are able to achieve a similar SED fit to LA19, including with the adoption of a smaller $0.14\,\mathrm{\mu m}$ grain size, it is noted that the smaller grain size is in line with older radiative transfer work of SU\,Aur by \citet{Akeson05}. The smaller grain size results in a smaller inner rim which now extends down to $0.13\,\mathrm{au}$ at a sublimation temperature of $2000\,\mathrm{K}$. This is within the uncertainties of values predicted by the temperature gradient modelling and is roughly consistent with older literature values of $0.18\pm0.04\,\mathrm{au}$ and $0.17\pm0.08\,\mathrm{au}$ by \citet{Akeson05} and \citet{Jeffers14} respectively. The flaring parameters $\alpha_{\mathrm{disk}}$ and $\beta_{\mathrm{disk}}$ were fixed such that $\alpha_{\mathrm{disk}} = \beta_{\mathrm{disk}} + 1$ and found to be $2.3$ and $1.3$ respectively, a more physical representation, than the values depicted in LA19. Similarly to LA19, a dusty disk wind is required in order to fit both the SED across the NIR and visibilities as shown in Figure\,\ref{fig:Vis_TORUS}. 
    
    The TORUS implementation of s dusty wind does not depend on the underlying launching mechanism, but just prescribes a geometry above and below the disk, which is populated by dust grains, where it can reprocess stellar radiation to contribute to the NIR excess. This was first put forward by \citet{Bans12} in the context of a magnetospherically driven disk wind. Figure\,\ref{fig:Vis_TORUS} highlights how a standard (no-wind) model cannot sufficiently fir the interferometric data, and how the addition of a dusty disk wind can provide a significantly better fit to the data. The maximum temperature of the dust in the wind is similar to the temperature of the dust at the sublimation temperature of the disk $1900\,\mathrm{K}$ and $2000\,\mathrm{K}$ respectively, as expected given the dust launches from close to to the sublimation rim. In Appendix\,\ref{AppB} each baseline is plotted in a separate panel in order to explore the chromatic (temperature) gradient of the data. The discrepancy in the gradient and level of some baselines can be explained by the simplified heating description within the radiative transfer model. TORUS produces a disk heated by reprocessed stellar radiation alone, with no internal disk heating such as viscous heating. We see from the specific temperature gradient modelling that we might expect some viscous heating within the disk.
    
    However, the implementation of the dusty disk wind in this scenario is not completely physical, owing to the high into-wind outflow rate of $1\times10^{-7}\mathrm{ M_{\odot}yr^{-1}}$ required. If one assumes the historically accepted outflow to accretion ratio of $0.1$, the resulting onto-star-accretion rate is greater than those typically found in T\,Tauri stars. In particular, this contrasts with the measured accretion rate of SU\,Aur \citep{Perez20}. However, the inflow to outflow ratio is subject of some discussion with recent works suggesting the ratio may be closer to unity in some situations \citep{Pascucci22}, particularly by invoking a magnetically driven winds originating from a dead-zone close to the sublimation radius. Based on our data we cannot differentiate such specific wind launching mechanisms. 
    In addition, the suggestion that SU\,Aur is undergoing a late stage infall event \citet{Ginski21} could be a potential explanation for such a high level of mass transport through the system. 
    
    Late infall events typically occur after the depletion of the protostellar envelope, when a large amount of material exterior to the disk falls inwards. The triggering mechanisms behind such events are not well understood as there are very few observational examples. From simulations it is thought that a large infall event can not only directly increase the accretion rate\citep{Dullemond19}, but also dramatically re-sculpt the disk, creating gaps, spiral arms and strong misalignments \citep{Kuffmeier21}. In the case of SU\,Aur a large infall of material is thought to have caused a strong disk warp between the inner and outer disk \citep{Ginski21}. It is also not unreasonable to assume this lead to an enhanced accretion rate, which would allow for the strong disks winds needed in the model presented here.

    The simplest way to directly compare the analytical methods employed in this work, is through the images produced. Figure\,\ref{fig:ImRec} shows a collage of the images produced by the geometric modelling, image reconstruction and radiative transfer methods. It is clearly shown the similarities between the images all with very similar inclinations and position angles with asymmetries oriented in the same direction. The asymmetries all appear to be the result of the high inclination of the disk causing the near-side inner rim to be shadowed by the flared outer disk. The lowest panel of Figure\,\ref{fig:ImRec} shows an image reconstructed from synthetic visibilties and closure phases obtained from the radiative transfer image. Care was taken to ensure the observables were matched in baseline length and position angle and detector mode, while the image reconstruction process was exactly that described in Section\,\ref{ImRec}. The remarkable similarity between the different images, in particular the real image reconstruction and the synthetic reconstruction, provides strong reinforcement that the models adopted are truly representative of the disk structure. 
    
    \subsection{Future Observations}
    
    Recent improvements at CHARA include the commissioning of the MYSTIC (Michigan Young STar Imager at CHARA) beam combiner \citep{Monnier18}, a 6-telescope K-band combiner capable of working in parallel with MIRC-X for simultaneous observations. The combination of MIRC-X and MYSTIC would allow us to investigate the temperature structure of the inner disk in greater detail, while also obtaining greater precision due to the higher sensitivity of MYSTIC. In addition, longer wavelength MIR interferometry with the MATTISE instrument at VLTI would potentially allow us to resolve the warped region of the disk, in order to confirm the nature of the geometry in this region for direct comparison to late infall event simulations.

\section{Conclusions}

This interferometric study of SU Aurigae has revealed the complex geometry and composition of the disk around SU\,Aurigae. We summarise our conclusions as follows:
    
    \begin{itemize}
        \item We reconstruct an interferometric image that confirms the inclined disk described in literature.  We see evidence for an asymmetry in the brightness distribution that can be explained by the exposure of the inner-rim on the far side of the disk and its obscuration on the near side due to inclination effects. 
        
        \item Our simple geometric model fits show that the circumstellar environment is best modelled as a Gaussian distribution with a disk of inclination $56.9\pm0.4^\circ$ along a minor axis position angle of $55.9.0\pm0.5^\circ$ and an FWHM of $1.52\pm0.01\,\mathrm{mas}$ ($0.239\pm0.002\,\mathrm{au}$). Such geometry is consistent with strong disk shadows observed in the outer disk originating from a disk warp. 
        
        \item We model the radial temperature profile of the inner disk and find a disk which extends down to $0.15\pm0.04\mathrm{au}$ where the temperature is equivalent to $2100\pm200\,K$ and decreases with an exponent of $0.62\pm0.02$. 
        
        \item A dusty disk wind scenario is still required to account for both the observed excess in the SED and the observed visibilities. The dusty disk wind scenario described here lifts material above the disk photosphere, thus exposing more dust grains to the higher temperatures close to the star responsible for the NIR excess. The high accretion rate required to reproduce the stellar-to-total flux ratio could be explained by a late infall event. 
        
        \item Our best-fit model (dusty disk wind model) suggests that the dust composition in the disk is dominated by small grains ($0.14\,\mathrm{\mu m}$) with a sublimation temperature of $2000$\,K. Introducing larger grains results in a worse fit to the SED shape and NIR excess. The disk is also shown to be highly flared ($9$\,au at $100$\,au).
        
    \end{itemize}

\begin{acknowledgements}
    We acknowledge support from an STFC studentship (No.\ 630008203) and an European Research Council Starting Grant (Grant Agreement No.\ 639889). MIRC-X received funding from the European Research Council (ERC) under the European Union's Horizon 2020 research and innovation programme (Grant No. 639889). JDM acknowledges funding for the development of MIRC-X (NASA-XRP NNX16AD43G, NSF-AST 1909165) and MYSTIC (NSF-ATI 1506540, NSF-AST 1909165).
    This research has made use of the VizieR catalogue access tool, CDS, Strasbourg, France. The original description of the VizieR service was published in \cite{Vizier}. 
\end{acknowledgements}

\bibliographystyle{aa}
\bibliography{REF}

\begin{thebibliography}{58}
\expandafter\ifx\csname natexlab\endcsname\relax\def\natexlab#1{#1}\fi

\bibitem[{{Abrahamyan} {et~al.}(2015){Abrahamyan}, {Mickaelian}, \&
  {Knyazyan}}]{Abrahamyan15}
{Abrahamyan}, H.~V., {Mickaelian}, A.~M., \& {Knyazyan}, A.~V. 2015, Astronomy
  and Computing, 10, 99

\bibitem[{{Akeson} {et~al.}(2005){Akeson}, {Walker}, {Wood}, {Eisner}, {Scire},
  {Penprase}, {Ciardi}, {van Belle}, {Whitney}, \& {Bjorkman}}]{Akeson05}
{Akeson}, R.~L., {Walker}, C.~H., {Wood}, K., {et~al.} 2005, \apj, 622, 440

\bibitem[{{Ammons} {et~al.}(2006){Ammons}, {Robinson}, {Strader}, {Laughlin},
  {Fischer}, \& {Wolf}}]{Ammons06}
{Ammons}, S.~M., {Robinson}, S.~E., {Strader}, J., {et~al.} 2006, \apj, 638,
  1004

\bibitem[{{Anderson} \& {Francis}(2012)}]{Anderson12}
{Anderson}, E. \& {Francis}, C. 2012, Astronomy Letters, 38, 331

\bibitem[{{Andrews} {et~al.}(2013){Andrews}, {Rosenfeld}, {Kraus}, \&
  {Wilner}}]{Andrews13}
{Andrews}, S.~M., {Rosenfeld}, K.~A., {Kraus}, A.~L., \& {Wilner}, D.~J. 2013,
  \apj, 771, 129

\bibitem[{{Anugu} {et~al.}(2020){Anugu}, {Le Bouquin}, {Monnier}, {Kraus},
  {Setterholm}, {Labdon}, {Davies}, {Lanthermann}, {Gardner}, {Ennis},
  {Johnson}, {ten Brummelaar}, {Schaefer}, \& {Sturmann}}]{Anugu20}
{Anugu}, N., {Le Bouquin}, J.-B., {Monnier}, J.~D., {et~al.} 2020, arXiv
  e-prints, arXiv:2007.12320

\bibitem[{{Bans} \& {K{\"o}nigl}(2012)}]{Bans12}
{Bans}, A. \& {K{\"o}nigl}, A. 2012, \apj, 758, 100

\bibitem[{{Baron} {et~al.}(2010){Baron}, {Monnier}, \& {Kloppenborg}}]{Baron10}
{Baron}, F., {Monnier}, J.~D., \& {Kloppenborg}, B. 2010, in Society of
  Photo-Optical Instrumentation Engineers (SPIE) Conference Series, Vol. 7734,
  Optical and Infrared Interferometry II, ed. W.~C. {Danchi}, F.~{Delplancke},
  \& J.~K. {Rajagopal}, 77342I

\bibitem[{{Bertout} {et~al.}(2007){Bertout}, {Siess}, \& {Cabrit}}]{Bertout07}
{Bertout}, C., {Siess}, L., \& {Cabrit}, S. 2007, \aap, 473, L21

\bibitem[{{Bianchi} {et~al.}(2011){Bianchi}, {Herald}, {Efremova}, {Girardi},
  {Zabot}, {Marigo}, {Conti}, \& {Shiao}}]{Bianchi11}
{Bianchi}, L., {Herald}, J., {Efremova}, B., {et~al.} 2011, \apss, 335, 161

\bibitem[{{Bonneau} {et~al.}(2006){Bonneau}, {Clausse}, {Delfosse}, {Mourard},
  {Cetre}, {Chelli}, {Cruzal{\`e}bes}, {Duvert}, \& {Zins}}]{Bonneau06}
{Bonneau}, D., {Clausse}, J.-M., {Delfosse}, X., {et~al.} 2006, \aap, 456, 789

\bibitem[{{Bonneau} {et~al.}(2011){Bonneau}, {Delfosse}, {Mourard}, {Lafrasse},
  {Mella}, {Cetre}, {Clausse}, \& {Zins}}]{Bonneau11}
{Bonneau}, D., {Delfosse}, X., {Mourard}, D., {et~al.} 2011, \aap, 535, A53

\bibitem[{{Bourg{\'e}s} {et~al.}(2014){Bourg{\'e}s}, {Lafrasse}, {Mella},
  {Chesneau}, {Bouquin}, {Duvert}, {Chelli}, \& {Delfosse}}]{Bourges14}
{Bourg{\'e}s}, L., {Lafrasse}, S., {Mella}, G., {et~al.} 2014, in Astronomical
  Data Analysis Software and Systems XXIII. Proceedings of a meeting held 29
  September - 3 October 2013 at Waikoloa Beach Marriott, Hawaii, USA. Edited by
  N. Manset and P. Forshay ASP conference series, vol. 485, 2014, p.223, Vol.
  485, 223

\bibitem[{{Castelli} \& {Kurucz}(2004)}]{Kurucz04}
{Castelli}, F. \& {Kurucz}, R.~L. 2004, ArXiv Astrophysics e-prints

\bibitem[{{Cutri} {et~al.}(2014){Cutri}, {Skrutskie}, {van Dyk}, {Beichman},
  {Carpenter}, {Chester}, {Cambresy}, {Evans}, {Fowler}, {Gizis}, {Howard},
  {Huchra}, {Jarrett}, {Kopan}, {Kirkpatrick}, {Light}, {Marsh}, {McCallon},
  {Schneider}, {Stiening}, {Sykes}, {Weinberg}, {Wheaton}, {Wheelock}, \&
  {Zacarias}}]{Cutri14}
{Cutri}, R.~M., {Skrutskie}, M.~F., {van Dyk}, S., {et~al.} 2014, VizieR Online
  Data Catalog, II/328

\bibitem[{{Davies} {et~al.}(2014){Davies}, {Gregory}, \& {Greaves}}]{Davies14}
{Davies}, C.~L., {Gregory}, S.~G., \& {Greaves}, J.~S. 2014, \mnras, 444, 1157

\bibitem[{{Davies} {et~al.}(2018){Davies}, {Kraus}, {Harries}, {Kreplin},
  {Monnier}, {Labdon}, {Kloppenborg}, {Acreman}, {Baron}, {Millan-Gabet},
  {Sturmann}, {Sturmann}, \& {Ten Brummelaar}}]{Davies18}
{Davies}, C.~L., {Kraus}, S., {Harries}, T.~J., {et~al.} 2018, \apj, 866, 23

\bibitem[{{Davies} {et~al.}(2022){Davies}, {Rich}, {Harries}, {Monnier},
  {Laws}, {Andrews}, {Bae}, {Wilner}, {Anugu}, {Ennis}, {Gardner}, {Kraus},
  {Labdon}, {Le Bouquin}, {Lanthermann}, {Schaefer}, {Setterholm}, {Ten
  Brummelaar}, \& {G-Lights Collaboration}}]{Davies22}
{Davies}, C.~L., {Rich}, E.~A., {Harries}, T.~J., {et~al.} 2022, \mnras, 511,
  2434

\bibitem[{{Draine}(2003)}]{Draine03}
{Draine}, B.~T. 2003, \araa, 41, 241

\bibitem[{{Dullemond} \& {Dominik}(2004)}]{Dullemond04}
{Dullemond}, C.~P. \& {Dominik}, C. 2004, \aap, 417, 159

\bibitem[{{Dullemond} {et~al.}(2019){Dullemond}, {K{\"u}ffmeier}, {Goicovic},
  {Fukagawa}, {Oehl}, \& {Kramer}}]{Dullemond19}
{Dullemond}, C.~P., {K{\"u}ffmeier}, M., {Goicovic}, F., {et~al.} 2019, \aap,
  628, A20

\bibitem[{{Eisner} \& {Hillenbrand}(2011)}]{Eisner11}
{Eisner}, J.~A. \& {Hillenbrand}, L.~A. 2011, \apj, 738, 9

\bibitem[{{Esplin} {et~al.}(2014){Esplin}, {Luhman}, \& {Mamajek}}]{Esplin14}
{Esplin}, T.~L., {Luhman}, K.~L., \& {Mamajek}, E.~E. 2014, \apj, 784, 126

\bibitem[{{Fitzpatrick}(1999)}]{Fitzpatrick99}
{Fitzpatrick}, E.~L. 1999, \pasp, 111, 63

\bibitem[{{Foreman-Mackey}(2016)}]{ForemanMackey16}
{Foreman-Mackey}, D. 2016, The Journal of Open Source Software, 1, 24

\bibitem[{{Ginski} {et~al.}(2021){Ginski}, {Facchini}, {Huang}, {Benisty},
  {Vaendel}, {Stapper}, {Dominik}, {Bae}, {M{\'e}nard}, {Muro-Arena},
  {Hogerheijde}, {McClure}, {van Holstein}, {Birnstiel}, {Boehler}, {Bohn},
  {Flock}, {Mamajek}, {Manara}, {Pinilla}, {Pinte}, \& {Ribas}}]{Ginski21}
{Ginski}, C., {Facchini}, S., {Huang}, J., {et~al.} 2021, \apjl, 908, L25

\bibitem[{{Harries}(2000)}]{Harries00}
{Harries}, T.~J. 2000, \mnras, 315, 722

\bibitem[{{Harries} {et~al.}(2019){Harries}, {Haworth}, {Acreman}, {Ali}, \&
  {Douglas}}]{Harries19}
{Harries}, T.~J., {Haworth}, T.~J., {Acreman}, D., {Ali}, A., \& {Douglas}, T.
  2019, Astronomy and Computing, 27, 63

\bibitem[{{Isella} \& {Natta}(2005)}]{Isella05}
{Isella}, A. \& {Natta}, A. 2005, \aap, 438, 899

\bibitem[{{Jang-Condell} \& {Turner}(2013)}]{Jang13}
{Jang-Condell}, H. \& {Turner}, N.~J. 2013, \apj, 772, 34

\bibitem[{{Jeffers} {et~al.}(2014){Jeffers}, {Min}, {Canovas}, {Rodenhuis}, \&
  {Keller}}]{Jeffers14}
{Jeffers}, S.~V., {Min}, M., {Canovas}, H., {Rodenhuis}, M., \& {Keller}, C.~U.
  2014, \aap, 561, A23

\bibitem[{{Kenyon} \& {Hartmann}(1987)}]{Kenyon87}
{Kenyon}, S.~J. \& {Hartmann}, L. 1987, \apj, 323, 714

\bibitem[{{Kluska} {et~al.}(2014){Kluska}, {Malbet}, {Berger}, {Baron},
  {Lazareff}, {Le Bouquin}, {Monnier}, {Soulez}, \& {Thi{\'e}baut}}]{Kluska14}
{Kluska}, J., {Malbet}, F., {Berger}, J.-P., {et~al.} 2014, \aap, 564, A80

\bibitem[{{K{\"o}nigl} \& {Salmeron}(2011)}]{Konigl11}
{K{\"o}nigl}, A. \& {Salmeron}, R. 2011, {The Effects of Large-Scale Magnetic
  Fields on Disk Formation and Evolution} ({University of Chicago Press}),
  283--352

\bibitem[{{Kraus} {et~al.}(2020){Kraus}, {Kreplin}, {Young}, {Bate}, {Monnier},
  {Harries}, {Avenhaus}, {Kluska}, {Laws}, {Rich}, {Willson}, {Aarnio},
  {Adams}, {Andrews}, {Anugu}, {Bae}, {ten Brummelaar}, {Calvet}, {Cur{\'e}},
  {Davies}, {Ennis}, {Espaillat}, {Gardner}, {Hartmann}, {Hinkley}, {Labdon},
  {Lanthermann}, {LeBouquin}, {Schaefer}, {Setterholm}, {Wilner}, \&
  {Zhu}}]{Kraus20}
{Kraus}, S., {Kreplin}, A., {Young}, A.~K., {et~al.} 2020, arXiv e-prints,
  arXiv:2004.01204

\bibitem[{{Kraus} {et~al.}(2018){Kraus}, {Monnier}, {Anugu}, {Le Bouquin},
  {Davies}, {Ennis}, {Labdon}, {Lanthermann}, {Setterholm}, \& {ten
  Brummelaar}}]{Kraus18}
{Kraus}, S., {Monnier}, J.~D., {Anugu}, N., {et~al.} 2018, in Society of
  Photo-Optical Instrumentation Engineers (SPIE) Conference Series, Vol. 10701,
  Optical and Infrared Interferometry and Imaging VI, 1070123

\bibitem[{{Kreplin} {et~al.}(2020){Kreplin}, {Kraus}, {Tambovtseva}, {Grinin},
  \& {Hone}}]{Kreplin20}
{Kreplin}, A., {Kraus}, S., {Tambovtseva}, L., {Grinin}, V., \& {Hone}, E.
  2020, \mnras, 492, 566

\bibitem[{{Kreplin} {et~al.}(2018){Kreplin}, {Tambovtseva}, {Grinin}, {Kraus},
  {Weigelt}, \& {Wang}}]{Kreplin18}
{Kreplin}, A., {Tambovtseva}, L., {Grinin}, V., {et~al.} 2018, \mnras, 476,
  4520

\bibitem[{{Kuffmeier} {et~al.}(2021){Kuffmeier}, {Dullemond}, {Reissl}, \&
  {Goicovic}}]{Kuffmeier21}
{Kuffmeier}, M., {Dullemond}, C.~P., {Reissl}, S., \& {Goicovic}, F.~G. 2021,
  \aap, 656, A161

\bibitem[{{Labdon} {et~al.}(2019){Labdon}, {Kraus}, {Davies}, {Kreplin},
  {Kluska}, {Harries}, {Monnier}, {ten Brummelaar}, {Baron}, {Millan-Gabet},
  {Kloppenborg}, {Eisner}, {Sturmann}, \& {Sturmann}}]{Labdon19}
{Labdon}, A., {Kraus}, S., {Davies}, C.~L., {et~al.} 2019, \aap, 627, A36

\bibitem[{{Labdon} {et~al.}(2021){Labdon}, {Kraus}, {Davies}, {Kreplin},
  {Monnier}, {Le Bouquin}, {Anugu}, {ten Brummelaar}, {Setterholm}, {Gardner},
  {Ennis}, {Lanthermann}, {Schaefer}, \& {Laws}}]{Labdon21}
{Labdon}, A., {Kraus}, S., {Davies}, C.~L., {et~al.} 2021, \aap, 646, A102

\bibitem[{{Lazareff} {et~al.}(2017){Lazareff}, {Berger}, {Kluska}, {Le
  Bouquin}, {Benisty}, {Malbet}, {Koen}, {Pinte}, {Thi}, {Absil}, {Baron},
  {Delboulb{\'e}}, {Duvert}, {Isella}, {Jocou}, {Juhasz}, {Kraus}, {Lachaume},
  {M{\'e}nard}, {Millan-Gabet}, {Monnier}, {Moulin}, {Perraut}, {Rochat},
  {Soulez}, {Tallon}, {Thi{\'e}baut}, {Traub}, \& {Zins}}]{Lazareff17}
{Lazareff}, B., {Berger}, J.-P., {Kluska}, J., {et~al.} 2017, \aap, 599, A85

\bibitem[{{Lucy}(1999)}]{Lucy99}
{Lucy}, L.~B. 1999, \aap, 344, 282

\bibitem[{{Mohanty} {et~al.}(2013){Mohanty}, {Greaves}, {Mortlock}, {Pascucci},
  {Scholz}, {Thompson}, {Apai}, {Lodato}, \& {Looper}}]{Mohanty13}
{Mohanty}, S., {Greaves}, J., {Mortlock}, D., {et~al.} 2013, \apj, 773, 168

\bibitem[{{Monnier} {et~al.}(2018){Monnier}, {Le Bouquin}, {Anugu}, {Kraus},
  {Setterholm}, {Ennis}, {Lanthermann}, {Jocou}, \& {ten
  Brummelaar}}]{Monnier18}
{Monnier}, J.~D., {Le Bouquin}, J.-B., {Anugu}, N., {et~al.} 2018, in Society
  of Photo-Optical Instrumentation Engineers (SPIE) Conference Series, Vol.
  10701, Optical and Infrared Interferometry and Imaging VI, ed. M.~J.
  {Creech-Eakman}, P.~G. {Tuthill}, \& A.~{M{\'e}rand}, 1070122

\bibitem[{{Morel} \& {Magnenat}(1978)}]{Morel78}
{Morel}, M. \& {Magnenat}, P. 1978, Astronomy and Astrophysics Supplement
  Series, 34, 477

\bibitem[{{Ochsenbein} {et~al.}(2000){Ochsenbein}, {Bauer}, \&
  {Marcout}}]{Vizier}
{Ochsenbein}, F., {Bauer}, P., \& {Marcout}, J. 2000, \aaps, 143, 23

\bibitem[{{Ofek}(2008)}]{Ofek08}
{Ofek}, E.~O. 2008, Publications of the Astronomical Society of the Pacific,
  120, 1128

\bibitem[{{Pascucci} {et~al.}(2022){Pascucci}, {Cabrit}, {Edwards}, {Gorti},
  {Gressel}, \& {Suzuki}}]{Pascucci22}
{Pascucci}, I., {Cabrit}, S., {Edwards}, S., {et~al.} 2022, arXiv e-prints,
  arXiv:2203.10068

\bibitem[{{P{\'e}rez} {et~al.}(2020){P{\'e}rez}, {Hales}, {Liu}, {Zhu},
  {Casassus}, {Williams}, {Zurlo}, {Cuello}, {Cieza}, \& {Principe}}]{Perez20}
{P{\'e}rez}, S., {Hales}, A., {Liu}, H.~B., {et~al.} 2020, \apj, 889, 59

\bibitem[{{Petrov} {et~al.}(2019){Petrov}, {Grankin}, {Gameiro}, {Artemenko},
  {Babina}, {Albuquerque}, {Djupvik}, {Gahm}, {Shenavrin}, {Irsmambetova},
  {Fernandez}, {Mkrtichian}, \& {Gorda}}]{Petrov19}
{Petrov}, P.~P., {Grankin}, K.~N., {Gameiro}, J.~F., {et~al.} 2019, \mnras,
  483, 132

\bibitem[{{Pollack} {et~al.}(1994){Pollack}, {Hollenbach}, {Beckwith},
  {Simonelli}, {Roush}, \& {Fong}}]{Pollack94}
{Pollack}, J.~B., {Hollenbach}, D., {Beckwith}, S., {et~al.} 1994, \apj, 421,
  615

\bibitem[{{Pringle}(1981)}]{Pringle81}
{Pringle}, J.~E. 1981, \araa, 19, 137

\bibitem[{{Renard} {et~al.}(2011){Renard}, {Thi{\'e}baut}, \&
  {Malbet}}]{Renard11}
{Renard}, S., {Thi{\'e}baut}, E., \& {Malbet}, F. 2011, \aap, 533, A64

\bibitem[{{R{\"o}ser} {et~al.}(2008){R{\"o}ser}, {Schilbach}, {Schwan},
  {Kharchenko}, {Piskunov}, \& {Scholz}}]{Roser08}
{R{\"o}ser}, S., {Schilbach}, E., {Schwan}, H., {et~al.} 2008, \aap, 488, 401

\bibitem[{{ten Brummelaar} {et~al.}(2005){ten Brummelaar}, {McAlister},
  {Ridgway}, {Bagnuolo}, {Turner}, {Sturmann}, {Sturmann}, {Berger}, {Ogden},
  {Cadman}, {Hartkopf}, {Hopper}, \& {Shure}}]{Brummelaar05}
{ten Brummelaar}, T.~A., {McAlister}, H.~A., {Ridgway}, S.~T., {et~al.} 2005,
  \apj, 628, 453

\bibitem[{{T{\'o}th} {et~al.}(2014){T{\'o}th}, {Marton}, {Zahorecz},
  {Bal{\'a}zs}, {Ueno}, {Tamura}, {Kawamura}, {Kiss}, \& {Kitamura}}]{Toth14}
{T{\'o}th}, L.~V., {Marton}, G., {Zahorecz}, S., {et~al.} 2014, Publications of
  the Astronomical Society of Japan, 66, 17

\bibitem[{{Unruh} {et~al.}(2004){Unruh}, {Donati}, {Oliveira}, {Collier
  Cameron}, {Catala}, {Henrichs}, {Johns- Krull}, {Foing}, {Hao}, {Cao},
  {Landstreet}, {Stempels}, {de Jong}, {Telting}, {Walton}, {Ehrenfreund},
  {Hatzes}, {Neff}, {B{\"o}hm}, {Simon}, {Kaper}, {Strassmeier}, \&
  {Granzer}}]{Unruh04}
{Unruh}, Y.~C., {Donati}, J.~F., {Oliveira}, J.~M., {et~al.} 2004, \mnras, 348,
  1301

\end{thebibliography}

\appendix

\section{Appendix A.} \label{AppA}
Table of photometry used in the SED fitting procedure.
    \begin{table*}
        \centering
        \caption{\label{table:Photometry} Photometric values used to construct the SED of SU\,Aur}
        \begin{tabular}{c c c} 
            \hline
            Wavelength [$\mathrm{\mu m}$] & Flux [Jy] & Reference \\
            \hline
            0.15&1.31E-04&\citet{Bianchi11}\\
            0.23&0.00219&\citet{Bianchi11}\\
            0.42&0.233&\citet{Ammons06}\\
            0.44&0.402&\citet{Anderson12}\\
            0.53&0.6&\citet{Ammons06}\\\
            0.69&1.32&\citet{Morel78}\\
            0.79&1.47&\citet{Davies14}\\
            0.88&1.75&\citet{Morel78}\\
            1.24&2.08&\citet{Roser08}\\
            1.25&2.12&\citet{Ofek08}\\
            1.63&2.47&\citet{Ofek08}\\
            2.17&2.71&\citet{Roser08}\\
            2.19&2.62&\citet{Ofek08}\\
            3.35&2.6&\citet{Cutri14}\\
            3.40&2.44&\citet{Bourges14}\\
            4.50&1.75&\citet{Esplin14}\\
            4.60&2.78&\citet{Cutri14}\\
            5.03&2.58&\citet{Bourges14}\\
            7.88&1.99&\citet{Esplin14}\\
            8.62&2.36&\citet{Abrahamyan15}\\
            11.57&2.83&\citet{Cutri14}\\
            11.60&3.52&\citet{Abrahamyan15}\\
            18.40&6.47&\citet{Abrahamyan15}\\
            22.11&9.24&\citet{Cutri14}\\
            23.90&12.8&\citet{Abrahamyan15}\\
            61.89&12.2&\citet{Abrahamyan15}\\
            65.04&9.89&\citet{Toth14}\\
            90.06&8.8&\citet{Toth14}\\
            140.10&10.2&\citet{Toth14}\\
            160.11&8.88&\citet{Toth14}\\
            849.86&0.074&\citet{Mohanty13}\\
            887.57&0.071&\citet{Andrews13}\\
            1300.90&0.03&\citet{Mohanty13}\\
            1333.33&0.0274&\citet{Andrews13}\\ [1ex]
            \hline
        \end{tabular}
    \end{table*}

\section{Appendix B.} \label{AppB}

\begin{figure*}
    \centering
    \includegraphics[scale=0.75]{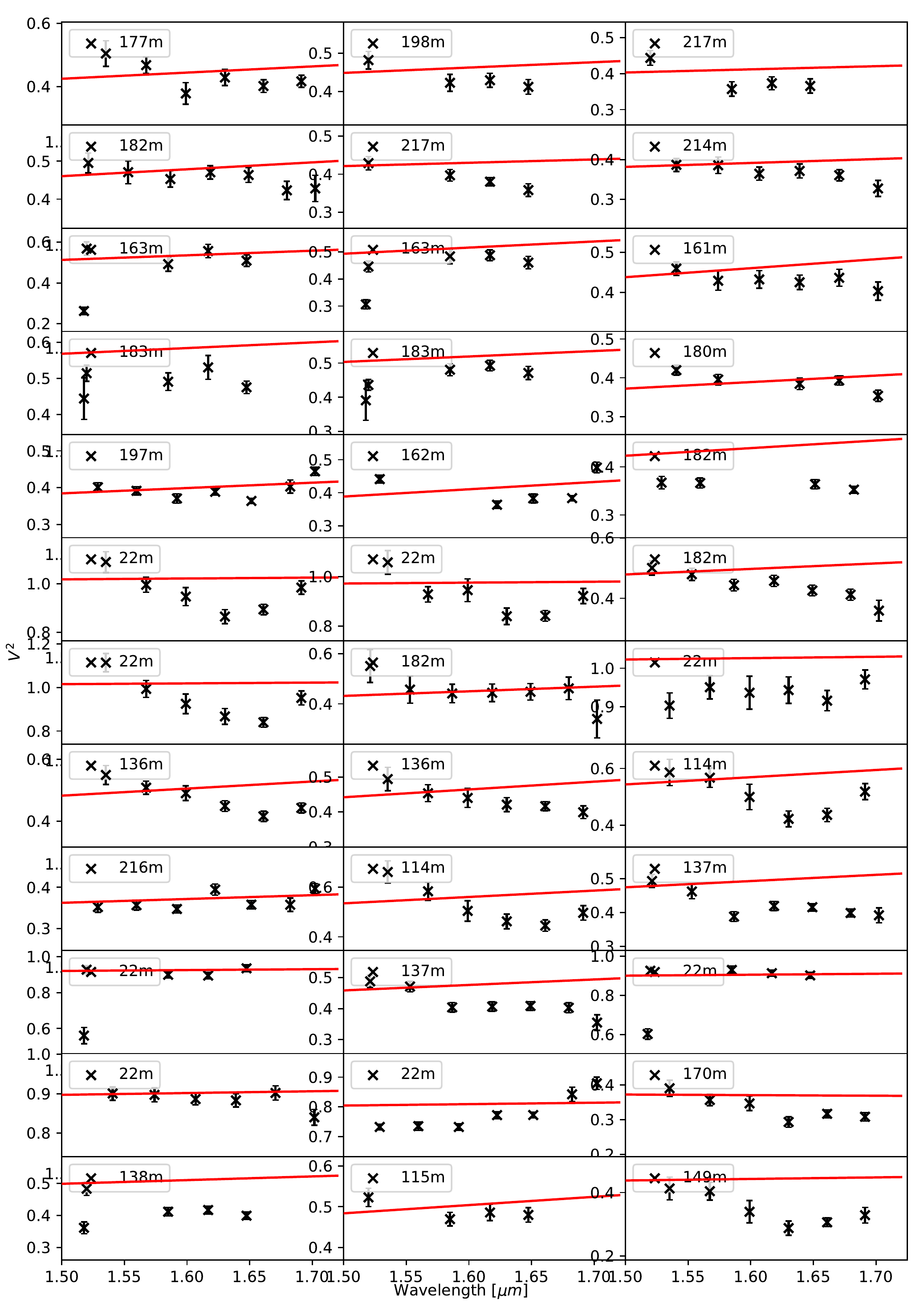}
    \caption{Squared visibilities split by baseline/pointing against wavelength. Black crosses are the measured visibilities. Red lines represent synthetic visibilities from the final TORUS radiate transfer model. The legend of each plot is the length of the individual baseline.}
    \label{fig:Vis_TORUS_BSplit_1}
\end{figure*}

\begin{figure*}
    \centering
    \includegraphics[scale=0.75]{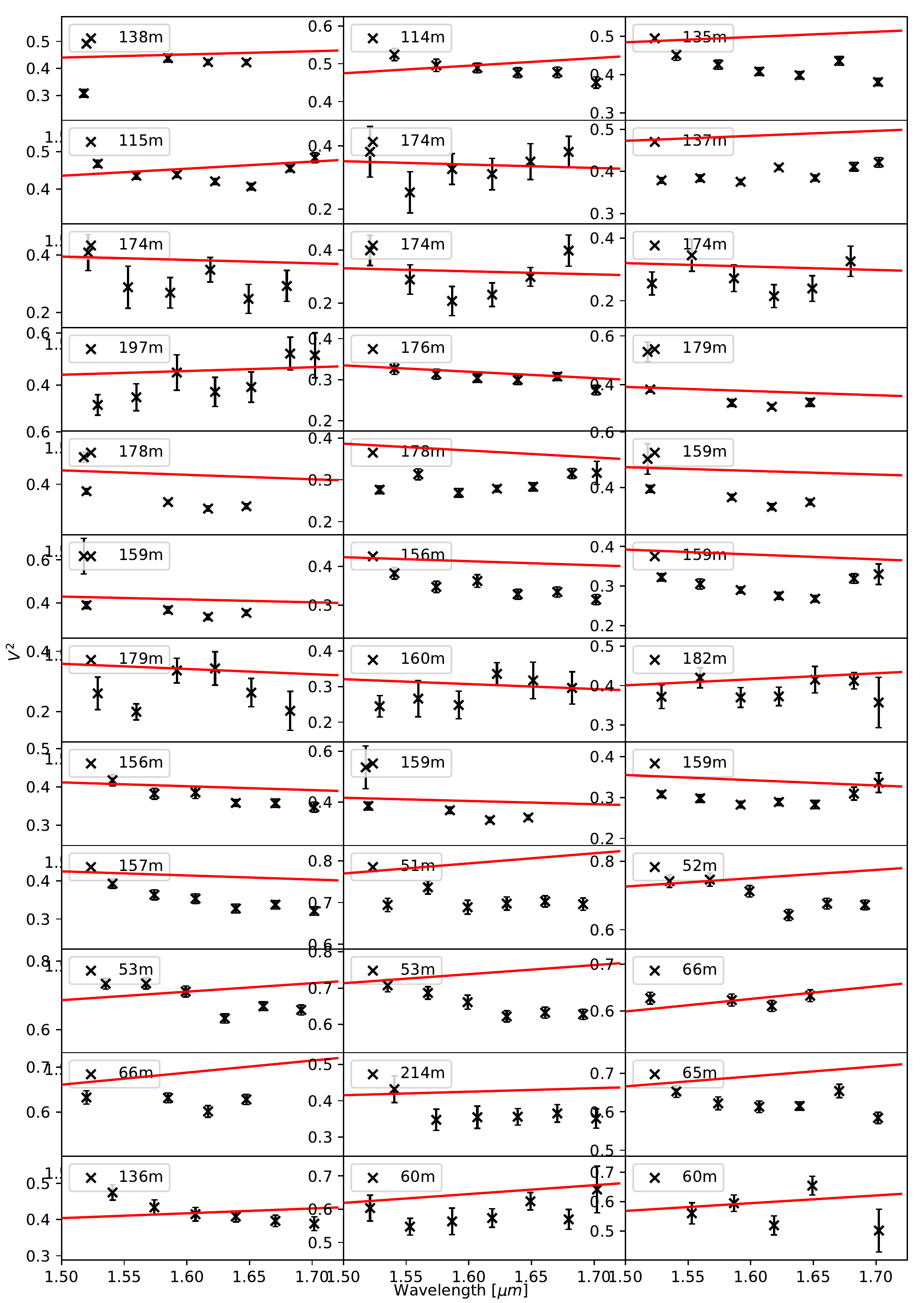}
    \caption{Figure\,\ref{fig:Vis_TORUS_BSplit_1} continued...}
\end{figure*}

\begin{figure*}
    \centering
    \includegraphics[scale=0.75]{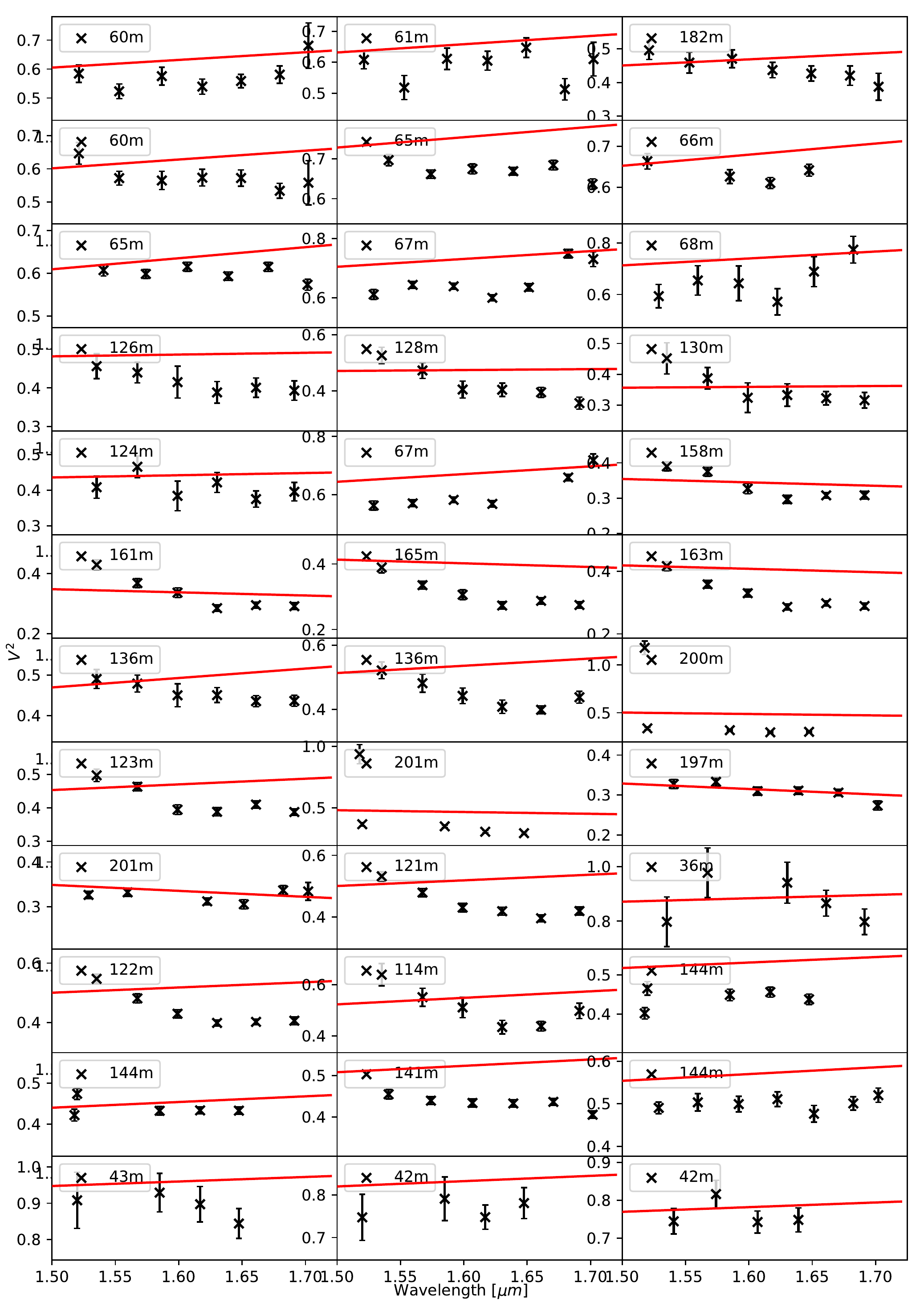}
    \caption{Figure\,\ref{fig:Vis_TORUS_BSplit_1} continued...}
\end{figure*}

\begin{figure*}
    \centering
    \includegraphics[scale=0.75]{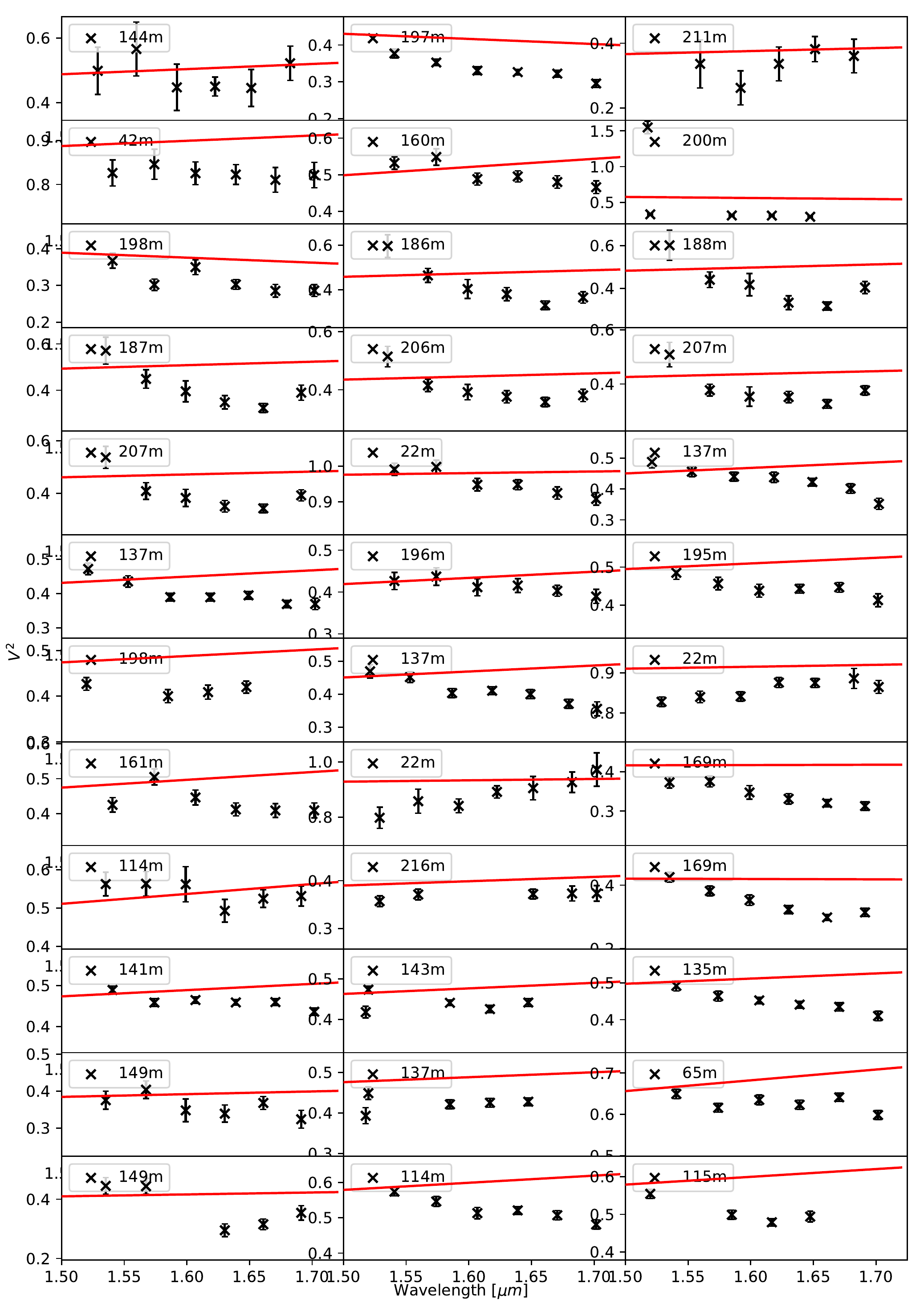}
    \caption{Figure\,\ref{fig:Vis_TORUS_BSplit_1} continued...}
\end{figure*}

\begin{figure*}
    \centering
    \includegraphics[scale=0.75]{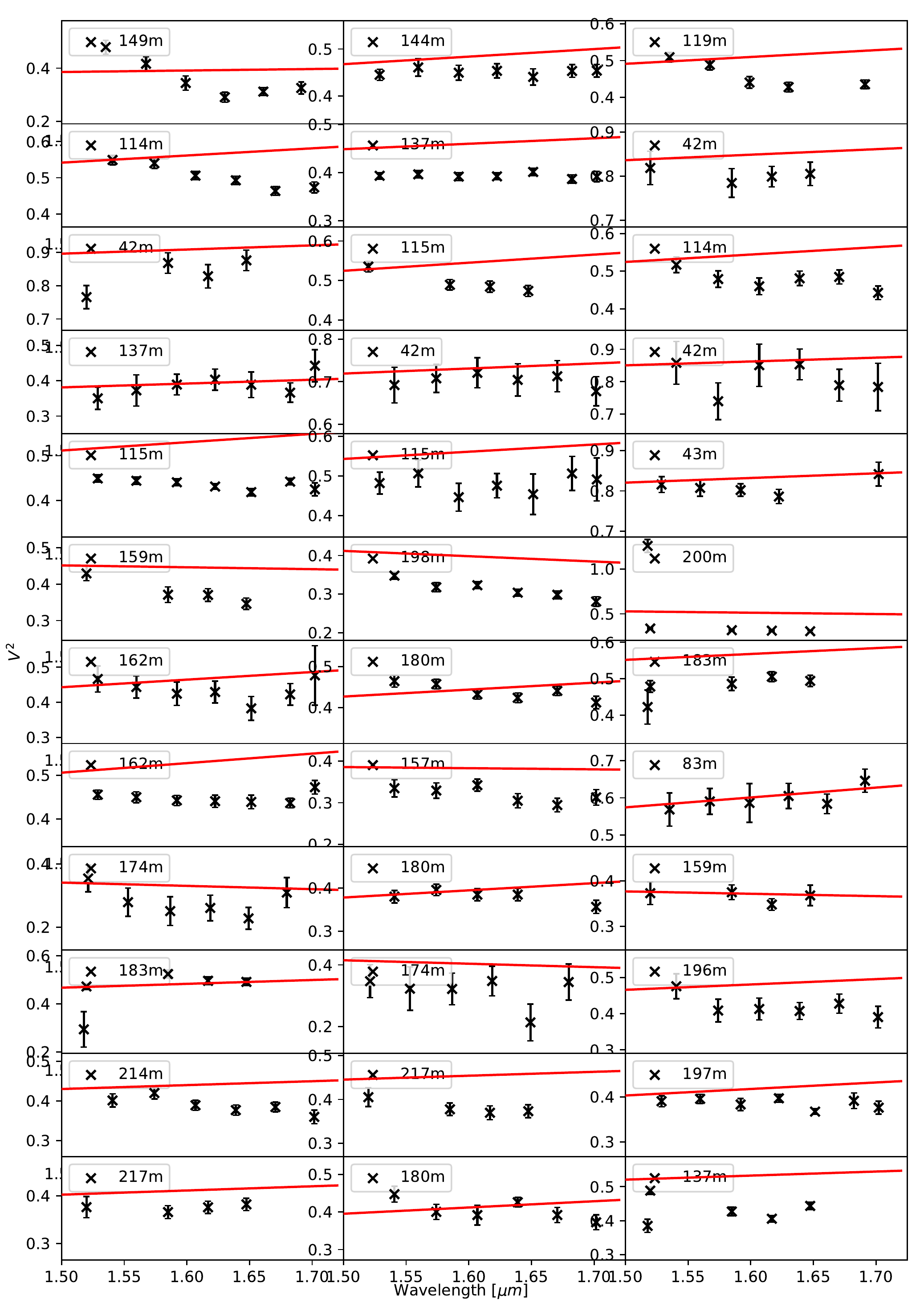}
    \caption{Figure\,\ref{fig:Vis_TORUS_BSplit_1} continued...}
\end{figure*}

\begin{figure*}
    \centering
    \includegraphics[scale=0.75]{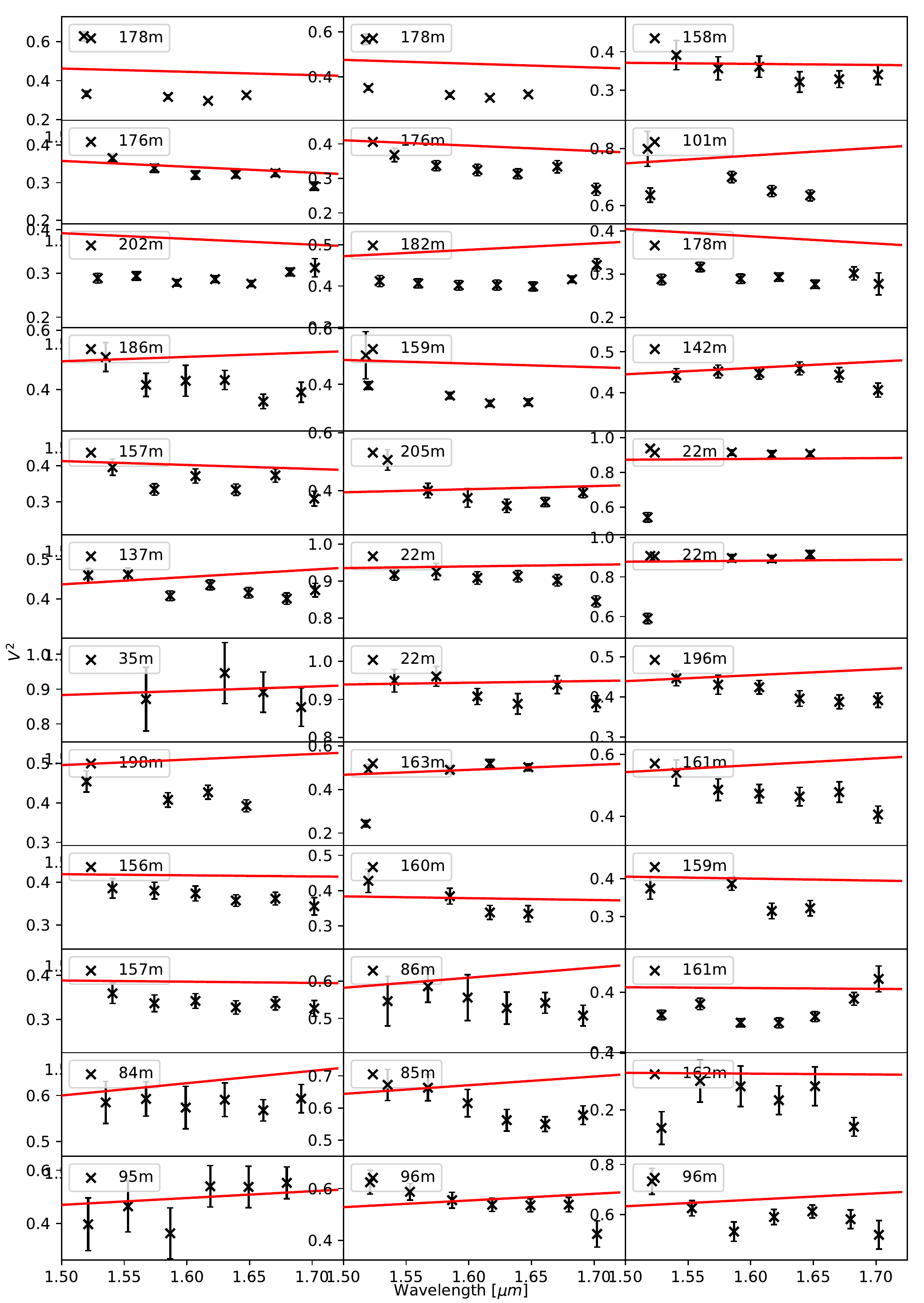}
    \caption{Figure\,\ref{fig:Vis_TORUS_BSplit_1} continued...}
\end{figure*}

\end{document}